\documentclass[aps,%
prd,%
twocolumn,%
nofootinbib,%
preprintnumbers,%
longbibliography,%
superscriptaddress]{revtex4-2}
\usepackage[utf8]{inputenc} 
\usepackage{hyperref} 
\usepackage{amsmath} 
\usepackage{amsfonts}
\usepackage{amsthm}
\usepackage{amssymb}
\usepackage{graphicx} 
\usepackage{slashed} 
\usepackage{color} 
\usepackage{subcaption} 
\usepackage{dcolumn} 
\usepackage[normalem]{ulem} 

\graphicspath{
  {./}
  {./figs/}
  {./figs/potential/}
  {./figs/cylindrical/}
}

\newcommand{\mr}[1]{\mathrm{#1}}
\newcommand{\mb}[1]{\mathbf{#1}}

\newcommand{\rmi}[1]{{\mbox{\scriptsize #1}}}

\newcommand{\lb}{\bar{\lambda}}

\newcommand{\kh}{\hat{\mb{k}}}
\newcommand{\lbtrap}{0.876088(1)} 
\newcommand{\phis}{\phi_{\rm f}}
\newcommand{\phib}{\phi_{\rm t}}
\newcommand{\Ms}{M_{\rm f}}
\newcommand{\Mb}{M_{\rm t}}
\newcommand{\bounce}{\phi_{0}}

\definecolor{OGcolour}{rgb}{0,0.5,0}

\newcommand{\Helsinki}{\affiliation{
    Department of Physics and Helsinki Institute of Physics,
    PL 64,
    FI-00014 University of Helsinki,
    Finland
}}

\newcommand{\Nottingham}{\affiliation{
    School of Physics and Astronomy,
    University of Nottingham,
    Nottingham NG7 2RD,
    U.K.
}}

\begin{document}

\title{Vacuum bubble collisions: from microphysics to gravitational waves}
\date{July 12, 2021} 

\author{Oliver Gould}
\email{oliver.gould@nottingham.ac.uk}
\Nottingham
\Helsinki

\author{Satumaaria Sukuvaara}
\email{satumaaria.sukuvaara@helsinki.fi}
\Helsinki

\author{David Weir}
\email{david.weir@helsinki.fi}
\Helsinki

\begin{abstract}
  We comprehensively study the effects of bubble wall thickness
  and speed on the gravitational wave emission spectrum of collisions
  of two vacuum bubbles.  We numerically simulate a large dynamical
  range, making use of symmetry to reduce the dimensionality.
  The high-frequency slope of the gravitational wave spectrum is shown to
  depend on the thickness of the bubble wall, becoming steeper for
  thick-wall bubbles, in agreement with recent fully 3+1 dimensional
  lattice simulations of many-bubble collisions.  This dependence is
  present, even for highly relativistic bubble wall collisions.
  We use the reduced dimensionality as an opportunity to
  investigate dynamical phenomena which may underlie the observed
  differences in the gravitational wave spectra. These phenomena include
  `trapping', which occurs most for thin-wall bubbles, and oscillations
  behind the bubble wall, which occur for thick-wall bubbles.  
\end{abstract}

\preprint{HIP-2021-4/TH}
\maketitle

\section{Introduction} \label{sec:introduction}

Observations of gravitational waves can provide a new probe of
fundamental physics.  In particular, the detection of a stochastic
gravitational wave background could provide some of the first
experimental data on the very early universe, long before
recombination.  Due to the universality of the gravitational coupling,
gravitational waves can also shed light on dark sectors, even if they
are not coupled directly to visible matter.

A first-order phase transition in the early universe would
produce a stochastic
gravitational wave background with characteristic broken power law spectral shape.
The shape is known to depend on several macroscopic thermodynamic
quantities, such as the temperature, strength and duration of the
phase transition as well as the speed at which bubble walls
expand~\cite{Caprini:2015zlo,Weir:2017wfa,Caprini:2019egz,Hindmarsh:2020hop}.
Gravitational wave detectors, such as the planned space-based
experiment LISA~\cite{Audley:2017drz,Caprini:2015zlo,Caprini:2019egz},
offer the exciting prospect of measuring a stochastic
gravitational wave background from a first-order phase transition, and
therefore of measuring these properties of the early universe.  From
this one can learn important information about the underlying particle
physics at the time of the first-order phase transition.

If the phase transition completes before much supercooling can take
place, the expanding bubble walls quickly reach a constant terminal
speed at which the vacuum pressure and the friction from the plasma
balance.  In this case sound waves propagating through the fluid
medium are thought to dominate the production of gravitational
waves~\cite{Hindmarsh:2013xza,Hindmarsh:2015qta,Hindmarsh:2016lnk,Hindmarsh:2017gnf,Hindmarsh:2019phv,Jinno:2020eqg}.
On the other hand, if there is sufficiently large supercooling, the
vacuum pressure may dominate over the friction from the plasma, and
the bubble wall will continue to accelerate until collision.  This is
referred to as a vacuum transition, and is the case we study here.
In this case the fluid dynamics of the plasma and its
interactions with the bubble wall are neglected.
Such a circumstance appears fine-tuned in Higgs transitions within
minimal electroweak extensions, due both to the relatively small
supercooling necessary for percolation to complete
in allowed regions of parameter space~\cite{Kehayias:2009tn,Chung:2012vg,Ellis:2018mja,Alves:2018jsw,Gould:2019qek,Kainulainen:2019kyp,Alves:2019igs}
and also the relatively large friction caused by the Higgs field's
interactions with Standard Model
particles~\cite{Bodeker:2009qy,Bodeker:2017cim,Ellis:2019oqb,Hoeche:2020rsg,Azatov:2020ufh}. On the
other hand, in dark sectors with relatively few degrees of
freedom~\cite{Baldes:2018emh,Fairbairn:2019xog,Baldes:2020kam,Huang:2020mso}, in
near-conformal extensions of the Standard
Model~\cite{Marzola:2017jzl,Ellis:2020nnr,Azatov:2020nbe}, and in certain QCD axion
models~\cite{Ghoshal:2020vud}, a large degree of supercooling is more
feasible.

Early studies of vacuum first-order phase transitions focused on the collisions of two
isolated bubbles, a system which has $\mathrm{O}(2,1)$ (hyperbolic)
symmetry~\cite{Hawking:1982ga,Wu:1984eda}, with the production of
black holes and the structure of the surrounding spacetime of
principal interest. There was also interest in the efficiency of
particle production~\cite{Kosowsky:1992vn}.

The gravitational wave (GW) power spectrum from colliding pairs of bubbles
was also studied, first in pairs of isolated
bubbles~\cite{Kosowsky:1991ua}. Although gravitational waves are not
produced by two perfectly isolated bubbles, due to their $\mathrm{O}(2,1)$ symmetry,
the finite duration of the phase transition breaks this symmetry and yields
sizeable gravitational wave production.
This study led directly to the development
of the `envelope approximation', where the bubble wall stress-energy
is approximated by a Dirac delta function which vanishes upon
collision~\cite{Kosowsky:1992rz,Kosowsky:1992vn}.  Furthermore, both
the scalar field simulations and the envelope approximation it
inspired produce a clear broken power law shape to the gravitational
wave power spectrum, with the peak frequency determined by the typical
bubble separation. In particular, the envelope approximation
gravitational wave power spectrum for many-bubble collisions
increases as $\omega^3$ at low
frequencies, and above the peak it decreases as $\omega^{-1}$,
where $\omega$ is the angular frequency~\cite{Huber:2008hg}.

However, for highly relativistic bubble collisions,
the large separation of scales between the Lorentz-contracted
bubble wall and the distance between bubbles meant that direct
numerical simulation of large numbers of colliding bubbles was
difficult, and so the envelope approximation became the main technique
used to study gravitational waves from first-order phase transitions~\cite{Huber:2008hg,Weir:2016tov}.
When direct numerical
simulations of gravitational waves from thermal phase transitions
became possible, it was found that long-lived sound waves were the
principal source of gravitational
waves~\cite{Hindmarsh:2013xza,Hindmarsh:2015qta}.

Nevertheless, for vacuum transitions the stress-energy was
perceived as being concentrated on the bubble wall. Therefore,
the use of the envelope approximation still seemed justified, until
direct numerical simulation showed rather a rather different spectral
shape~\cite{Child:2012qg,Cutting:2018tjt}. The simulations found
a steeper high-frequency power law of $\omega^{-1.5}$, and additional
high-frequency gravitational wave production due to the dynamics of
the field about the true vacuum.
  
Further new insights have been gained from simulations of vacuum
  transitions in recent years, as computational capabilities have
improved and simulation volumes have
increased~\cite{Jinno:2019bxw,Lewicki:2019gmv,Cutting:2020nla}.
These have revealed a surprisingly rich parameter space due to
the nonlinear phenomena present during and after the collisions
of two bubble walls.

As a result of these new computer simulations, and renewed interest in
phase transitions more generally, vacuum phase transitions have become
the subject of a recent debate.  In Ref.~\cite{Jinno:2019bxw} a phenomenon
was studied whereby the kinetic energy released in a bubble collision
causes the field to bounce back to the metastable false vacuum.
\footnote{
    This phenomenon had previously been described in Refs.~\cite{Hawking:1982ga,Wu:1984eda,Watkins:1991zt}, though not explored specifically.
}

The term {\em trapping} was coined to describe this phenomenon, which was observed to occur for thin-wall bubbles, but not for thick-wall bubbles.
For the collision of two planar walls, trapping was shown to occur permanently, with a region of space unable to escape to the true vacuum.
This qualitative difference between the collisions of thick and thin
wall bubbles motivated the possibility of an observable effect in the
gravitational wave spectrum.

Furthermore, Ref.~\cite{Jinno:2019bxw} showed that the effect of
trapping also depends on the velocity of the bubble walls at
collision. Many direct numerical simulations of bubble collisions in
vacuum transitions have been carried out in three dimensions. In three
dimensions computational limitations on lattice sizes significantly
limit the dynamic range for bubbles to accelerate to large gamma
factors; a system with reduced dimensionality would allow more
extensive studies. However, the geometry of (1+1)-dimensional planar
bubble walls studied in Ref.~\cite{Jinno:2019bxw} is physically very
different to that of colliding spherical bubbles in
(3+1)-dimensions. Working with the reduced dimensionality of the
hyperbolic two-bubble collision system will allow us to explore the
parameter space of trapping more thoroughly while retaining the
three-dimensional geometry.

Perhaps for this very reason, the hyperbolic two-bubble system
  has seen some recent interest.  Ref.~\cite{Lewicki:2019gmv} studied
the GW spectrum of two-bubble collisions, for two sets of parameter
choices, one producing thinner and the other thicker bubble walls.
They found that the GW spectra were very similar for their two
benchmark points, which led them to conclude that there was no
difference between the GW spectra of collisions of thick- and thin-wall
bubbles.  Ref.~\cite{Cutting:2020nla} simulated collisions of many
vacuum bubbles for four different bubble wall thicknesses.  By
contrast, they found a strong dependence of the GW spectrum on the
bubble wall width.  In particular, there it was shown that the
gravitational wave power spectrum high-frequency power-law
$\omega^{-b}$ with index $b$
was steeper for thick-wall bubbles than for thin-wall bubbles,
varying from $b = 1.36 \pm 0.05$ to $b = 2.25 \pm 0.18$.

To resolve this debate requires a thorough study of the parameter
space of vacuum bubble collisions.  In the simplest model, the real
scalar theory, there are two parameters: the bubble wall thickness,
and the Lorentz factor of the bubble wall at collision.  Ideally, one
would perform fully 3+1 dimensional simulations of many-bubble
collisions.  However, such simulations use a significant amount of
computer resources for a single run.  In this paper, we study
two-bubble collisions, for which one can reduce the
the problem to (1+1)-dimensions in hyperbolic coordinates, and
comprehensively study the parameter space of the minimal real scalar
theory.

Properly understanding the spectral shape of vacuum bubble
collisions will allow us to infer properties of the phase transition,
if a stochastic gravitational wave background is detected. It is
therefore important to study both the power law dependence and the
nonlinear dynamics that result.
Today, the spectral shape remains a significant source of
uncertainty~\cite{NANOGrav:2021flc}.

In Section~\ref{sec:bubble_dynamics} we introduce our scalar field model,
the symmetries of the problem, and the geometry in which we study
bubble collisions. Next, in Section~\ref{sec:gravitational_waves}, we
discuss the methods we use to compute the gravitational wave power
spectrum and extract the spectral shape. Our results are presented in
Section~\ref{sec:results}, with discussion following in
Section~\ref{sec:conclusions}.

\section{Bubble dynamics} \label{sec:bubble_dynamics}

The basic principles of vacuum bubble nucleation and collision can be
studied with a single-component scalar field $\phi$, for which the
potential has a tree-level barrier. We therefore have the action
\begin{equation} \label{eq:action}
S[\phi] =  \int d^4 x \left(\partial_\mu \phi \partial^\mu \phi - \frac{1}{2}m^2 \phi^2 + \frac{\delta}{3} \phi^3 - \frac{\lambda}{4} \phi^4\right),
\end{equation}
with $m$ the mass parameter, and $\delta$ and $\lambda$ the cubic and
quartic couplings.  This is the simplest renormalisable field theory
with a first-order phase transition; the simpler $Z_2$-symmetric theory has only a
second-order phase transition.  More complicated theories with
additional field content may lead to qualitatively different
dynamics~\cite{Copeland:1996jz,Saffin:1997kr,Copeland:1999ua,Johnson:2003ti,Jinno:2019bxw,Lewicki:2020jiv,Lewicki:2020azd,Di:2020ivg}.
 
In principle $\phi$ may be a scalar field in a fundamental UV theory, or simply an effective operator describing the order parameter of the transition.%
\footnote{
For example, the gauge-invariant condensate $\langle H^\dagger H\rangle$, which distinguishes between the two phases of a Higgs-like phase transition, is a real scalar.
}
We consider potential parameters such that there is a first-order phase transition from a metastable \emph{false vacuum} at $\langle \phi \rangle = 0$ to a stable \emph{true vacuum} at $\langle \phi \rangle \neq 0$.
Note that any linear (tadpole) term in the Lagrangian can be removed by a shifting of the field origin.
The parameters should be understood to be the effective parameters of the low-energy theory which describes physics at the length-scales relevant for bubble nucleation.

Bubble nucleation may proceed either via quantum mechanical tunnelling
or a thermal over-barrier transition.  We assume the transition to
take place via quantum mechanical tunnelling, and hence that the
temperature is much smaller than the inverse of the bubble radius at
nucleation~\cite{Linde:1981zj}.
In this case the nucleation process
effectively happens in vacuum, and the bubble has $\mathrm{O}(4)$
symmetry.  At higher temperatures, for which there is a thermal
over-barrier transition, the nucleated bubble instead has
$\mathrm{O}(3)$ symmetry.

In either case, after nucleation the bubble is highly occupied
and hence semiclassical.  In this paper, we will assume that the
smooth classical field equations resulting from Eq.~\eqref{eq:action}
provide a sufficiently accurate description for the time evolution.
Corrections to this description, arising from the effect of thermal or
quantum mechanical fluctuations, can be incorporated by adding
stochastic fluctuation and dissipation terms to the equations of
motion, or to the initial conditions.  We further assume that the
field undergoing the transition does not interact sufficiently
strongly with other fields to affect its dynamics.

The time evolution of both $\mathrm{O}(3)$ and $\mathrm{O}(4)$ bubbles in
vacuum was considered in Ref.~\cite{Lewicki:2019gmv}, where it was found
that at late times no significant difference between the two was observed.
Note however that for $\mathrm{O}(3)$ bubbles, the presence of the thermal
bath may significantly affect the time evolution equations, except perhaps
in the case of thermal runaways~\cite{Bodeker:2009qy}.

We also assume a flat Minkowski background spacetime, so
that for example the transition is not so slow and strong that
the nucleation of bubbles causes inflation by virtue of the vacuum
energy released~\cite{Guth:1982pn}.

\begin{figure}
  \centering
  \includegraphics[width=0.45\textwidth]{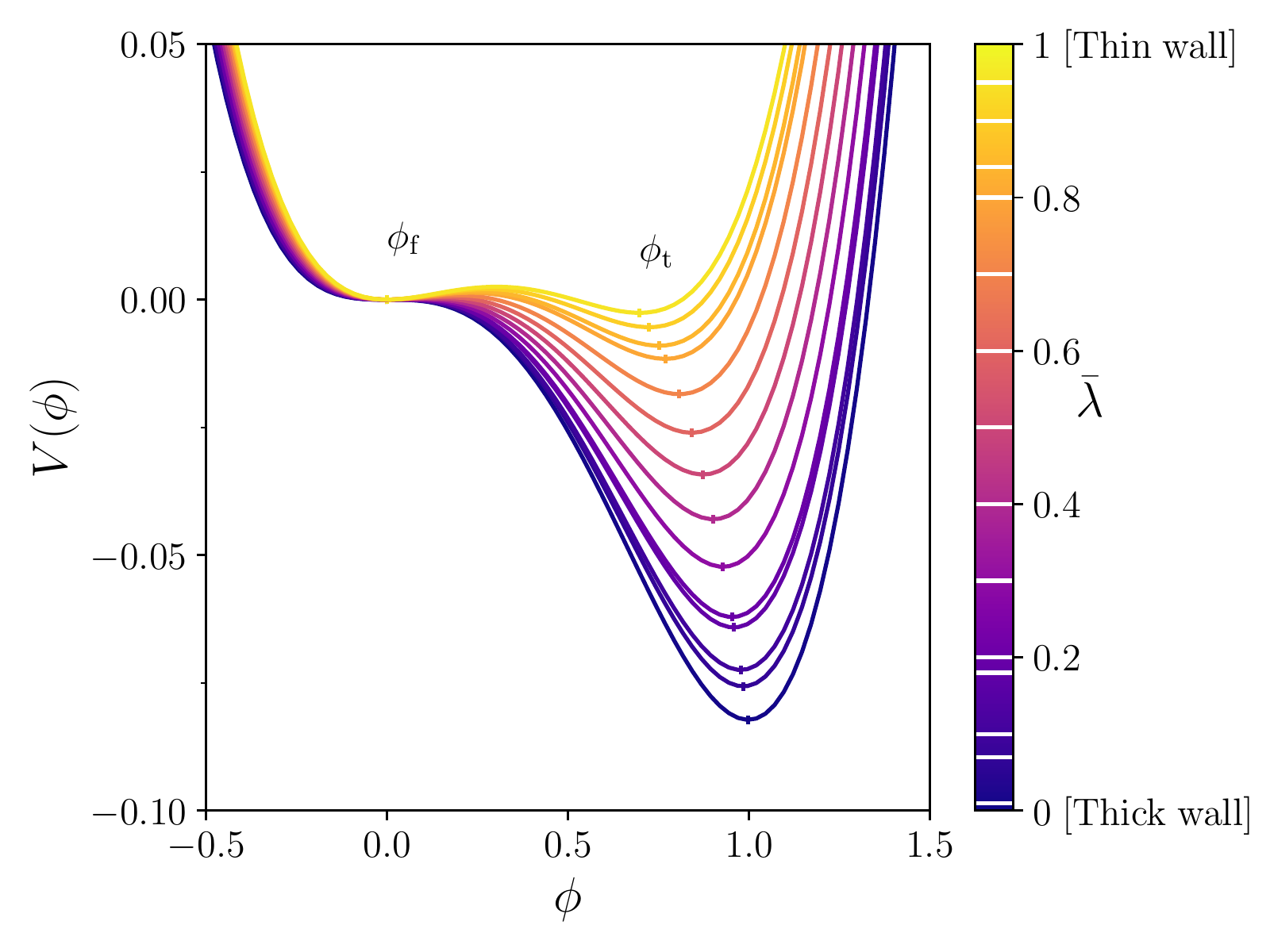}
  \caption{\label{fig:potential} Sketch of the potential
    (\ref{eq:potential}) used in this paper. The values of
    $\overline{\lambda}$ at which we simulate are indicated by
    horizontal white lines in the colour bar. The
    false and true vacua, $\phis$ and $\phib$
    respectively, are indicated with small vertical lines in the
    potential curves. See also Fig.~1 in Ref.~\cite{Cutting:2020nla},
    in which the curves are normalised by the vacuum energy difference
    $\Delta V = V(\phib) - V(\phis)$.}
\end{figure}

The parametric dependence of the classical theory can be simplified by the following transformation:
\begin{equation}
x^\mu \to \frac{\sqrt{\lambda}}{\delta} x^\mu, \qquad \phi \to \frac{\delta}{\lambda} \phi
\end{equation}
Under this transformation the action transforms to
\begin{align} \label{eq:action_scaled}
S[\phi] = \frac{1}{\lambda}\int d^4 x \left( \partial_\mu \phi \partial^\mu \phi - \frac{\lambda m^2}{2\delta^2}\phi^2 + \frac{1}{3} \phi^3 - \frac{1}{4} \phi^4 \right).
\end{align}
Thus the classical dynamics after nucleation only depends nontrivially on the combination,
\begin{equation}
\lb \equiv \frac{m^2}{m_c^2} = \frac{9\lambda m^2}{2\delta^2},
\end{equation}
where $m_c$ is the critical mass,
at which point the two phases are degenerate in energy.
In this parameterisation, the potential
energy density reads,
\begin{equation} \label{eq:potential}
V(\phi) = \frac{\lb}{9}\phi^2 - \frac{1}{3} \phi^3 + \frac{1}{4} \phi^4.
\end{equation}
This parameterisation was introduced in Ref.~\cite{Enqvist:1991xw}, and has been used since in, for example, Ref.~\cite{Cutting:2020nla}.
For convenience, the relation to some other conventions is given in Appendix~\ref{appendix:conventions}.
The minima for this potential are located at
\begin{equation}
\phis = 0, \qquad \phib = \frac{1}{2} \left(1+\sqrt{1-\frac{8}{9}\lb }\right),
\end{equation}
and we will focus on the case where these are a metastable false vacuum and a stable true vacuum respectively, i.e.\ where $V(\phis)>V(\phib)$ and both are minima.
For $\lb>1$ the extremum at $\phi=0$ is the global minimum, and at $\lb=1$ it is degenerate with the other minimum at $\phi \neq 0$.
At and below $\lb=0$ there is no longer a barrier between the two vacua,
and hence there can be no first-order phase transition;
starting from $\phi=0$, spinodal decomposition will occur for such values of $\lb$.
A first-order phase transition from $\phis$ to $\phib$
may take place for $\lb \in (0,1)$.
The thick- and thin-wall limits are given by
\begin{equation}
\text{thick: }\lb \to 0_+, \qquad \text{thin: }\lb \to 1_-.
\end{equation}
We plot the potential used
in Figure~\ref{fig:potential}.

The critical bubble solves the bounce equations~\cite{Langer:1967ax,Coleman:1977py},
\begin{equation} \label{eq:tunnelling}
\frac{d^2\bounce}{d \rho^2} + \frac{3}{\rho}\frac{d\bounce}{d\rho} - \frac{d V}{d\bounce} = 0,
\end{equation}
with boundary conditions such that $\bounce\to \phis$ as $\rho\to\infty$
and $d\bounce/d\rho =0$ at the origin.
In this paper we solve the bounce equation using the {\tt CosmoTransitions} code~\cite{Wainwright:2011kj}.
In all cases tunnelling takes place from the false vacuum through the potential barrier, ending somewhat short of the true vacuum~\cite{Coleman:1977py}.
For thin-wall bubbles, tunnelling takes place almost up to $\phib$, whereas for thick-wall bubbles the tunnelling trajectory falls far short.
As we will later show, this difference has important consequences for the dynamics of the phase transition.

Note that, for bubble nucleation to take place at a given cosmological time,
it must be that the bubble nucleation rate is at least as fast as the Hubble expansion.
This implies the following relation between $\lambda$, $\lb$ and the Hubble rate $H$ in units of the particle mass~\cite{Enqvist:1991xw,Anderson:1991zb}
\begin{equation} \label{eq:cosmological_nucleation}
\frac{1}{\lambda}\bar{S}(\lb) \sim 4\log\left(\frac{m}{H}\right),
\end{equation}
where $\bar{S}(\lb) \equiv \lambda\ S(m,\delta,\lambda)$
is the scaled action of the critical bubble, which
ranges from 0 to infinity as $\lb$ ranges from 0 to 1
(see Eq.~\eqref{eq:action_scaled}).
As a consequence of
Eq.~\eqref{eq:cosmological_nucleation}, for a given $m/H$, more
weakly coupled particles (smaller $\lambda$) will nucleate with
thicker wall bubbles (smaller $\lb$).

As long as both $m$ and the energy density of the universe are far
sub-Planckian, it will be that $m/H \gg 1$.  Therefore bubble
nucleation will take place when the (unscaled) action of the critical
bubble is large.  As long as the rate of change of the bubble
nucleation rate is not so large as to counteract the $m/H \gg 1$
hierarchy, the average distance between nucleated bubbles $R_*$ will
be large compared with their initial radius $R_0$; for details see for
example Refs.~\cite{Hawking:1982ga,Enqvist:1991xw,Anderson:1991zb}.
In this case the bubbles have a long time to expand before collision,
and hence, under constant acceleration due to the vacuum energy
difference between phases, they will reach highly relativistic
velocities.

\subsection{Symmetries}

The critical bubble is invariant under a Euclidean $\mathrm{O}(4)$ symmetry
about its centre.
Its time evolution is determined by the Wick rotation of the bounce equation, and hence, after nucleation, it has an $\mathrm{O}(3,1)$ symmetry,
\begin{equation}
\phi(x) = \bounce(-t^2+x^2+y^2+z^2).
\end{equation}
However, in performing the Wick rotation, a choice for the initial time slice is made, which would appear to break the $\mathrm{O}(3,1)$ down to $\mathrm{O}(3)$.
The question arises though, as to what physically breaks this symmetry.
The answer, as made clear in Refs.~\cite{Garriga:2012qp,Garriga:2013pga}, is that an observer is required to break this symmetry, as all inertial observers will see bubbles preferentially nucleated at rest.
Therefore, in the absence of an observer
the evolution of a vacuum bubble has $\mathrm{O}(3,1)$ and not just $\mathrm{O}(3)$ symmetry.

In the presence of a second critical bubble, nucleated in a spacelike separated region, the line joining their centres defines a preferred direction.
We may define this line as being along the $z$ axis and choose a Lorentz frame in which the bubbles are nucleated simultaneously.
As a result of this preferred direction half of the symmetries are broken.
The remaining unbroken symmetry generators are the rotations about the $z$ axis, $\mb{J}_z$, and the boosts in the $x$ and $y$ directions, $\mb{K}_x$ and $\mb{K}_y$, which together form the generators of $\mathrm{O}(2,1)$,
\begin{equation}
[\mb{J}_z,\mb{K}_x] = \mb{K}_y, \
[\mb{J}_z,\mb{K}_y] = -\mb{K}_x, \
[\mb{K}_x,\mb{K}_y] = -\mb{J}_z,
\end{equation}
with all other commutators zero.  Just as in the case of a single
bubble, the initial conditions defined at some initial time, $t=0$,
break the boost symmetries, reducing the symmetry group down to the
$\mathrm{O}(2)$ group generated by $\mb{J}_z$.  However, due to the
bubbles being spacelike separated, the notion of simultaneous
nucleation is contingent upon an inertial observer.  Thus, in the
absence of an observer, the evolution of two vacuum bubbles has
$\mathrm{O}(2,1)$ and not just $\mathrm{O}(2)$ symmetry.

To make manifest the $\mathrm{O}(2,1)$ symmetry, one can use hyperbolic coordinates $(s,\psi,\theta,z)$, defined in two patches in terms of the Cartesian coordinates $(t,x,y,z)$.
Following Ref.~\cite{Lewicki:2019gmv}, we label the patches by $+$ and $-$ for the complementary regions $t^2\geq x^2+y^2$ and $t^2\leq x^2+y^2$ respectively.
In region $+$ the coordinates and metric, $dl^2$, are given by:
\begin{align} \label{eq:coords_begin}
t =& s \cosh\psi,\ x = s \sinh\psi\cos\theta,\ y = s \sinh\psi\sin\theta, \\
dl^2 &= ds^2 - s^2 d\psi^2 - s^2 \sinh^2(\psi)d\theta^2 - dz^2,
\end{align}
and in the complementary region $-$, they are:
\begin{align}
t =& s \sinh\psi,\ x = s \cosh\psi\cos\theta,\ y = s \cosh\psi\sin\theta, \\
dl^2 &= - ds^2 + s^2 d\psi^2 - s^2 \cosh^2(\psi)d\theta^2 - dz^2,
\label{eq:coords_end}
\end{align}
where we have adopted the mostly minus signature.

The coordinates $\psi$ and $\theta$ are transformed nontrivially under $\mathrm{O}(2,1)$ transformations, whereas $s$ and $z$ are left unchanged.
As a consequence the field describing the two-bubble system is independent of $\psi$ and $\theta$.
The equations of motion are
\begin{equation} \label{eq:eom}
 \pm \frac{\partial^2 \phi_\pm}{\partial s^2} \pm \frac{2}{s} \frac{\partial \phi_\pm}{\partial s} - \frac{\partial^2 \phi_\pm}{\partial z^2} + \frac{d V}{d\phi_\pm} = 0,
\end{equation}
where $+$ and $-$ in $\pm$ refer to the regions $t^2>x^2+y^2$ and $t^2<x^2+y^2$ respectively.
This is a hyperbolic partial differential equation (PDE) for $t^2>x^2+y^2$ and an elliptic PDE for $t^2<x^2+y^2$.

There is an important caveat to this $\mathrm{O}(2,1)$ symmetry.  The
bounce, the most likely path between minima, has $\mathrm{O}(4)$
symmetry.  However, the weight of any single, specific field
configuration in the path integral is zero.  When considering the
process of bubble nucleation, one must sum over the phase space in the
vicinity of the bounce, giving the so-called fluctuation prefactor in
the rate of bubble nucleation~\cite{Callan:1977pt}.  The addition of
statistical fluctuations to the background field breaks the
$\mathrm{O}(4)$ symmetry of the bounce by a small amount, and in their
evolution some fluctuations may be exponentially
amplified~\cite{Braden:2014cra,Braden:2015vza,Bond:2015zfa}.  Once the
fluctuations have grown sufficiently large and nonlinear, the symmetry
of the original background field configuration is completely broken.

In our analysis, we choose to utilise the $\mathrm{O}(2,1)$ symmetry of the two-bubble system without statistical fluctuations.
The consequent reduction in computational effort allows us to study a much greater dynamical range than would be possible if we were to study the full 3+1 dimensional problem.
In particular, this allows us to study significantly larger collision velocities than were possible in the 3+1 dimensional studies of Refs.~\cite{Bond:2015zfa,Cutting:2018tjt}.
However, in our setup we cannot study the growth of
small symmetry-breaking
fluctuations and the eventual breakdown of
the approximate $\mathrm{O}(2,1)$ symmetry.
Cause for optimism can nevertheless be found in the 3+1 dimensional simulations of Ref.~\cite{Bond:2015zfa},
in which the effect of small symmetry-breaking fluctuations was investigated.
There two-bubble collisions were studied, one with thin and the other with thick walls, equivalent to $\lb\approx 0.941$ and $\lb\approx 0.0223$ respectively.
The thin-wall case showed exponential growth of fluctuations partially resulting from the trapping phenomenon, with significant deviation from the $\mathrm{O}(2,1)$ symmetry only after approximately twice the time taken for the bubbles to accelerate and collide.
We will stop our simulations at or before this time.
Further, for their thick-wall bubble collision Ref.~\cite{Bond:2015zfa} found that the symmetry-breaking fluctuations did not grow significantly even at late times.

\subsection{Solving the equations of motion} \label{sec:solving_eoms}

Here we briefly describe how we set initial conditions and solve the field equations of motion, Eq.~\eqref{eq:eom}.
In general, our approach utilises a rectangular lattice in $(z,s)$, with derivatives approximated by finite differences.
Tests of this approximation, and of our numerical implementation~\cite{sukuvaara_satumaaria_2021_5127538} are collected in Appendix~\ref{appendix:tests}.

Two bubble configurations are initialised at $s=0$, solutions of the bounce equations.
Their origins are located a distance $d$ apart, with $d$ chosen
such that the two bubbles will collide with a given Lorentz factor,
\begin{equation} \label{eq:gamma_d_R0}
\gamma = \frac{d}{R_0}.
\end{equation}
Here $R_0$ is the bubble radius, defined to be the point at which $\bounce(R_0)=\tfrac{1}{2}\bounce(0)$.
For highly relativistic bubble collisions, the bubbles are initially far apart, though for small enough $\gamma$, their exponential tails may overlap.
This overlap issue is handled as in Ref.~\cite{Cutting:2018tjt}.
The definition of the Lorentz factor given in Eq.~\eqref{eq:gamma_d_R0} is based on the speed of movement of the field profile, or more specifically of the point with field value $\phi=\phi_0(R_0)$.

An alternative definition of $\gamma$, based upon the Lorentz contraction of the bubble wall, was put forward in Ref.~\cite{Cutting:2020nla},
\begin{align} \label{eq:gamma_d_Rin_Rout}
\gamma_{\rm alt} &= \frac{R_{\rm out}-R_{\rm in}}{\sqrt{R_{\rm out}^2+s_{\rm col}^2}-\sqrt{R_{\rm in}^2+s_{\rm col}^2}}, \\
s_{\rm col}^2 &= \left(\frac{d}{2}\right)^2-R_0^2, \label{eq:s_col}
\end{align}
written in terms of the inner and outer bubble radii, defined as $\phi_0(R_{\rm in})= 0.731 \phi_0(0)$ and $\phi_0(R_{\rm out})= 0.269 \phi_0(0)$ respectively.
The differences between these two definitions of the Lorentz factor are largest for thick-wall bubbles, and vary from less than 0.1\% for $\lb=0.9$ to as much as 5\% for $\lb=0.01$.

The equations of motion are solved separately in the two regions referred to in Eq.~\eqref{eq:eom}.
In the timelike $+$ region, $t^2>x^2+y^2$, the bubbles collide and the (hyperbolic) equations of motion must be solved numerically.
To do so, we have adopted a leap-frog algorithm, which converges quadratically as the discretisation scales, $dz$ and $ds$, are taken to zero.
Given the presence in Eq.~\eqref{eq:eom} of both first and second order derivatives in $s$, our algorithm takes the form of a Crank-Nicolson algorithm~\cite{crank1947practical,Figueroa:2020rrl}.
The explicit discrete equations are collected in Appendix~\ref{appendix:discrete_eom}.
From the initial conditions at $s=0$, this algorithm calculates the field at positions $ds, 2ds, 3ds, \dots$ and the field momentum at positions $ds/2, 3ds/2, 5ds/2, \dots$.
To describe the initial half-step of the momentum field with the same accuracy as the following steps, we have used the trick of splitting it up into many smaller steps with size $\ll ds$.

In the spacelike $-$ region, $t^2<x^2+y^2$, the bubbles never meet and the (elliptic) equation of motion \eqref{eq:eom} is equivalent to the tunnelling equation, Eq.~\eqref{eq:tunnelling}.
Thus, the solution in this region can be written simply in terms of the bounce solution~\cite{Lewicki:2019gmv}
\begin{align}
\phi_-(s,z) &= \bounce\Big(\sqrt{s^2 + \left(z-d/2\right)^2}\ \Big) \nonumber \\
 &\qquad + \bounce\Big(\sqrt{s^2 + \left(z+d/2\right)^2}\ \Big).
\end{align}

\subsection{Linear modes} \label{sec:linear_modes}

In general, in both regions, the equation of motion must be solved
numerically.  However, in the $+$ region, for small oscillations
around one of the minima, $\phi_0 \in \{\phis,\phib\}$, we can expand
Eq. \eqref{eq:eom} to linear order in $\varphi = \phi-\phi_0$,
\begin{equation} \label{eq:eom_linear}
 \left(\frac{\partial^2 }{\partial s^2} + \frac{2}{s} \frac{\partial }{\partial s} - \frac{\partial^2 }{\partial z^2} + M^2 \right) (\phi-\phi_0)= 0.
\end{equation}
For the scaled potential, Eq.~\eqref{eq:potential}, the scaled masses, $M$,
around the false and true vacua are,
\begin{equation}
\Ms^2 = \frac{2\lb}{9}, \qquad \Mb^2 = \frac{1}{2} \left(1-\frac{8}{9}\lb + \sqrt{1-\frac{8}{9} \lb }\right).
\end{equation}
The original dimensionful masses are attained from these scaled
masses by multiplication by $\delta^2/\lambda$, so that
$\delta^2\Ms^2/\lambda=m^2$.
Note that in the thick-wall limit $\lb \to 0$, $\Ms \to
  0_+$ and $\Mb \to 1_-$, while in the thin-wall case $\lb \to
  1$, $\Ms$ and $\Mb$ both tend to $\sqrt{2}/3$.

The solution to the linearised equation of motion can be found by
Fourier transforming Eq. (\ref{eq:eom_linear}) with respect to $z$
and then noting that the resulting equation is a Bessel equation.
The general solution to Eq.~\eqref{eq:eom_linear} is
\begin{equation}
 \phi = \phi_0 + \int \frac{dk}{2\pi}\left(\alpha(k) f_k(s,z) +\alpha^*(k) f_k^*(s,z)\right), \label{eq:linear_solution} 
\end{equation}
where the wave modes are,
\begin{equation}
f_k(s,z) = \frac{\mathrm{e}^{-i\sqrt{M^2+k^2}s+ikz}}{s}.
\end{equation}
These modes describe the free-particle or linear-wave solutions about the minima, with dispersion relation,
\begin{equation} \label{eq:dispersion_relation}
\omega^2 = M^2 + k^2.
\end{equation}
Relaxing the dispersion relation, the modes form a complete basis with which to expand the field.
If the field is well described by a superposition of linear excitations about one of the minima, the dominant modes in the expansion will satisfy Eq.~\eqref{eq:dispersion_relation}.

On the lattice, we adopt a discrete mode expansion which is orthogonal and approaches Eq.~\eqref{eq:linear_solution} in the continuum limit.
The details of our numerical implementation~\cite{sukuvaara_satumaaria_2021_5127538} are given in Appendix~\ref{appendix:fourier}.

\section{Gravitational waves} \label{sec:gravitational_waves}

Gravitational waves are sourced by shear stresses, by the transverse, traceless part of the energy-momentum tensor.
In highly symmetric systems, such as those with spherical $\mathrm{O}(3)$ symmetry, the net gravitational wave production is zero.
In fact, it was shown in Ref.~\cite{Wu:1984eda} that this is also the case for the $\mathrm{O}(2,1)$-symmetric collision of two vacuum bubbles.
As gravitational waves are sourced locally, but the symmetry is a global property, their absence can be understood as due to precise cancellations between the gravitational waves produced by different regions.

In a cosmological first-order phase transition, the $\mathrm{O}(2,1)$ symmetry of two-bubble collisions is broken by their coming into contact with additional bubbles, which eventually fill the universe with the new phase and end the transition.
For our two-bubble collisions, this process can be modelled by cutting off the collision in an $\mathrm{O}(2,1)$-breaking way.
We follow Refs.~\cite{Kosowsky:1991ua,Lewicki:2019gmv} in choosing a constant time slice $t=t_{\rm c}$ to end the simulation of the collision, thereby breaking the two boost symmetries.
The duration of the phase transition is determined by the interplay of the cosmological expansion and the rate of change of the bubble nucleation rate~\cite{Enqvist:1991xw}.
It is found to scale linearly with the average
bubble separation, $t_{\rm c} \propto d$, where the constant
of proportionality is independent of $\lb$.
We will assume the completion of
the phase transition to be after the two-bubble collision that we will focus on,
in which case the Lorentz factor at collision is independent of the precise choice of $t_c$.
While alternative choices for modelling the end of the transition will lead to different gravitational wave spectra, we will be interested in the dependence of the spectrum on the parameters $\lb$ and $\gamma$, 
and such dependence may be revealed using any reasonable, fixed cutoff model.

We will work in the linearised gravity approximation, meaning that we consider only small metric fluctuations about the background Minkowski space, and ignore gravitational backreaction.
This means, in particular, that we do not include the effect of the
false vacuum inflating, which becomes relevant for very slow transitions,
and neither are we able to study black hole formation.
Our analysis is however fully (special) relativistic, which is necessary as the bubble walls and subsequent scalar field oscillations move with relativistic speeds. 

We are interested in the gravitational wave power radiated to infinity.
This can be determined in terms of the Fourier transform of the energy-momentum tensor,
\begin{equation}
T^{ij}(\omega,\mb{k}) = \frac{1}{2\pi} \int dt\ \mr{e}^{i\omega t} \int d^3 x\  \mr{e}^{- i\mb{k}\cdot \mb{x}} T^{ij}(\mb{x},t),
\end{equation}
where $\omega$ is the angular frequency and $\mb{k}$ is the momentum vector.
Only the components with null four-momentum, $\mb{k}=\omega \kh$ where $\kh$ is a unit vector, contribute to the gravitational wave spectrum.

The power radiated as gravitational waves from a localised source in a direction $\kh$ is given by the Weinberg formula~\cite{Weinberg:1972kfs},
\begin{align}
\frac{dE_{\rm GW}}{d\Omega d\log(\omega)} &= 2 G \omega^3 \Lambda_{ij,lm}(\kh ) T^{ij*}(\omega,\mb{k})T^{lm}(\omega,\mb{k}), \label{eq:weinberg_formula} \\
\Lambda_{ij,lm}(\hat{\mb{k}}) &= \delta_{il}\delta_{jm} - 2 \kh_j \kh_m \delta_{il} + \frac{1}{2}\kh_i\kh_j\kh_l\kh_m \nonumber \\
&\qquad - \frac{1}{2}\delta_{ij}\delta_{lm} + \frac{1}{2}\delta_{ij}\kh_l\kh_m + \frac{1}{2}\delta_{lm}\kh_i \kh_j.
\end{align}
Note that this formula has been derived in the far-field approximation (or
wave-zone), i.e.\ at distances from the source, $r$, much larger than
the wavelengths under consideration, $r \gg 1/\omega$, 
much larger than the size of the source, $r \gg R_{\rm source}$,
and $r \gg \omega R_{\rm source}^2$.
We will however follow previous
literature~\cite{Kosowsky:1991ua,Kosowsky:1992vn,Lewicki:2019gmv,Lewicki:2020jiv}
in using the formula down to its breaking point, $\omega r \sim
1$ and $r/R_{\rm source}\sim 1$.
We justify this by noting that we are chiefly interested in the
differences between the gravitational wave spectrum of collisions at
different $\lb$ and $\gamma$, rather than their absolute gravitational
wave spectrum.  Further, by focusing on two-bubble collisions, we are
anyway unable to describe the low-frequency physics of a system of
many colliding bubbles.  Thus we focus on the high-frequency tail of
the gravitational wave spectrum, between the peak and the microscopic
mass scale.  These wavelengths are smaller than the distance between
bubbles and hence should be well captured by two-bubble collisions,
and for them the far-field approximation is better justified.
Going beyond the far-field approximation can be achieved either at the
expense of more difficult numerical integrals, or by dynamically evolving
the metric fluctuations.

The translation of the Weinberg formula into hyperbolic coordinates has been given in Eqs.~(32) and (A1)-(A8) of Ref.~\cite{Kosowsky:1991ua}, which we have verified and utilised.\footnote{The same equations are also given in Eqs.~(20)-(21) of Ref.~\cite{Lewicki:2019gmv}, though they differ there by an overall factor of 1/4.} The result is a set of four integrals over the coordinates $(s,\psi,\theta,z)$, which we perform numerically.
As discussed in Ref.~\cite{Kosowsky:1991ua}, the integrals over $s$, $\psi$ and $z$ take the form of a pair of double integrals, rather than a full triple integral, which reduces significantly the numerical effort.

To implement the $\mathrm{O}(2,1)$-breaking end of the two-bubble collision, the gravitational wave integrals are multiplied by a cutoff function $C$.
The cutoff function used has the same form as in Ref.~\cite{Kosowsky:1991ua}, having an exponentially decreasing factor after a certain cutoff time $t_{\rm c}$,
\begin{align}
C(t=us)=\begin{cases}
            1 & t \leq t_{\rm c} \\
            e^{-(t-t_{\rm c})^2/t_0^2} & t>t_{\rm c}
        \end{cases}
\end{align}
where the coordinates $s$, $t$ and $u$ are those given in Eqs.~\eqref{eq:coords_begin} to \eqref{eq:coords_end}.
In our final simulations, we have chosen $t_{\rm c}=0.9\ s_{\rm max}$, $t_0=0.25\ (s_{\rm max} - t_{\rm c})$ and $s_{\rm max}=1.2\ d$.

The numerical integrations were performed using the trapezium rule, which converges quadratically to the continuum limit as the discretisation scales are decreased.
This therefore matches the accuracy of the leap-frog algorithm used to solve the scalar equations of motion.
Further details and tests of the numerical implementation~\cite{sukuvaara_satumaaria_2021_5127538} are collected in Appendix~\ref{appendix:tests}.

The gravitational wave spectrum produced by two-bubble collisions has a global maximum peak at $\omega_{\rmi{peak}} \approx \pi/d$ and power-law tails~\cite{Kosowsky:1991ua}.
The same is true for the gravitational waves produced by the many-bubble collisions of a full phase transition, with the peak position at $\omega_{\rmi{peak}} \approx \pi/R_*$, where $R_*$ is the mean separation of bubbles at nucleation~\cite{Cutting:2018tjt}.
In both cases, the spectrum shows additional structure at frequencies of order the mass of the scalar particle $\omega\sim M \ll \omega_{\rmi{peak}}$, though with a much lower amplitude than the main peak.  For gravitational wave experiments with limited
sensitivity, the vicinity of the main peak of the spectrum is of
primary interest.

The gravitational wave spectrum in the vicinity of the peak can be fit with the function~\cite{Cutting:2020nla},
\begin{equation} \label{eq:fit_fn}
\frac{d\Omega_{\rm fit}}{d\log(\omega)} = \tilde{\Omega}_{\rm GW} \frac{(a+b)\ \omega^a \tilde{\omega}^b}{a\ \omega^{a+b} + b\ \tilde{\omega}^{a+b}},
\end{equation}
where $a$, $b$, $\tilde{\omega}$ and $\tilde{\Omega}_{\rm GW}$ are the fit parameters.
The parameters $a$ and $b$ correspond to the low-frequency $\omega^a$ and high-frequency $\omega^{-b}$ power laws respectively, while $\tilde{k}$ and $\tilde{\Omega}_{\rm GW}$ approximately correspond to the peak position and amplitude.
Note that here high frequencies correspond to those in the window $\omega_{\rmi{peak}} \ll \omega \ll M$.

Fits were performed by minimising the sum of squared residuals, the default behaviour of the {\tt scipy.optimize.curve\_fit} function in {\tt SciPy 1.5.3}.
The fit is performed on a restricted range of data, satisfying $\omega < \omega_{\rm cut}$, where $\omega_{\rm cut} = \mathrm{min}(\Ms,\Mb,10\pi/d)$, to avoid both mass-scale contributions and numerical artefacts.
This choice was further motivated by the desire not to cut off the peak for the smallest values of the Lorentz factor.
We have verified that varying $\omega_{\rm cut}$ by a factor of 2 has no
significant effect on the fit results at $\gamma\gtrsim 4$, because
points in the vicinity of the peak dominate the sum of squared
residuals. Therefore, we do not anticipate substantial bias from
  the low- or high-frequency power laws.

The low-frequency power law for the gravitational wave spectrum can be
argued to be $\omega^3$ based on causality~\cite{Caprini:2009fx}.
Within our current framework the same result can be arrived at
as follows.  For a localised source of gravitational waves, such as we
consider, the small-frequency limit of the Fourier-transformed
energy-momentum tensor is a finite constant.  Assuming this
  constant is nonzero, from Eq.~\eqref{eq:weinberg_formula} we can
see that the low-frequency power law for the gravitational wave
spectrum is $\omega^3$.  We therefore set $a=3$ in
Eq.~\eqref{eq:fit_fn}.

\begin{figure*}[p]
  \centering
  \begin{subfigure}{\textwidth}
    \includegraphics[width=0.75\textwidth]{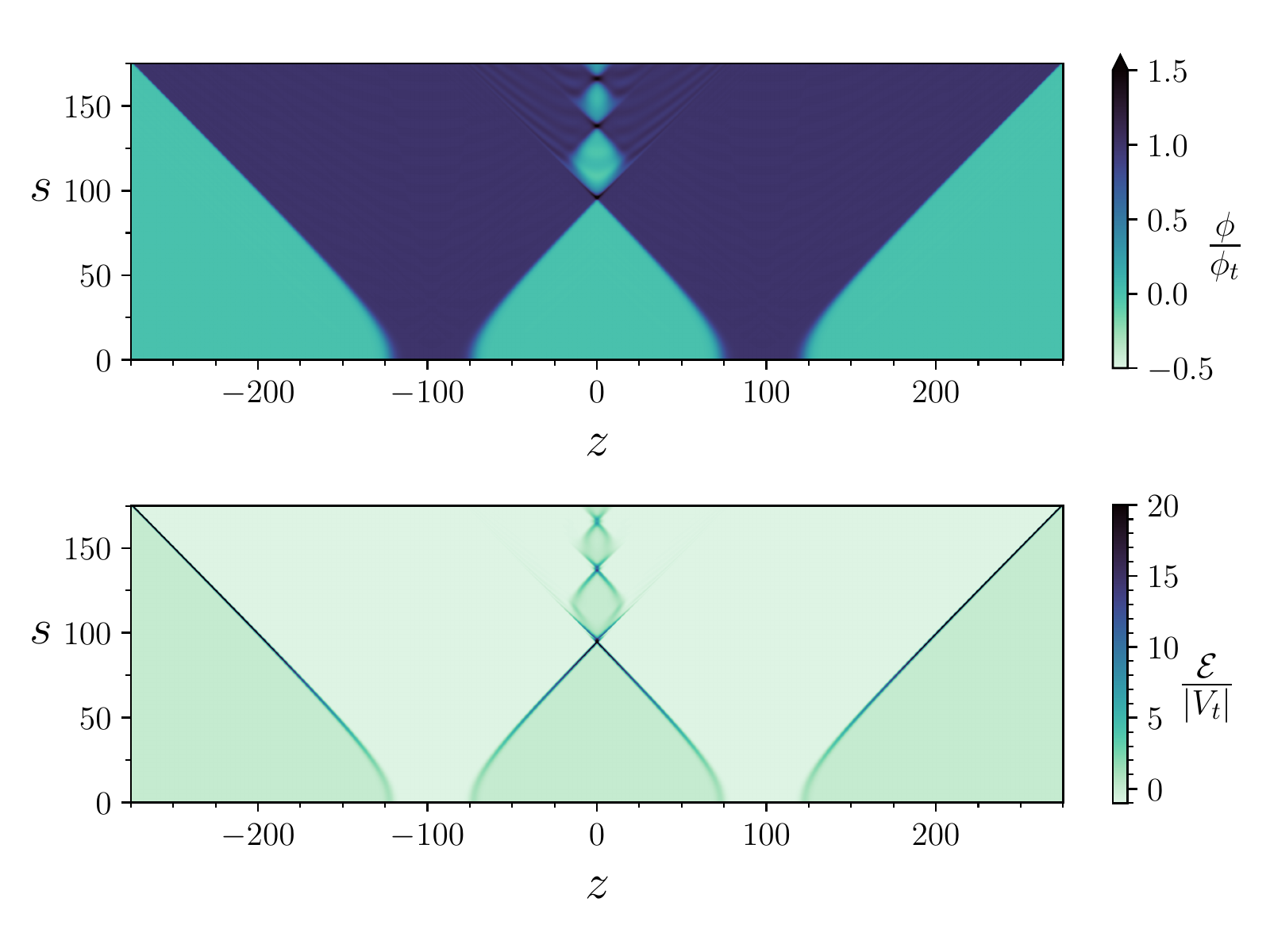}
    \caption{
      Thin-wall bubble collision, at $(\lb,\gamma)=(0.9,4)$.
      \label{fig:field_thin}
    }
  \end{subfigure}
    \begin{subfigure}{\textwidth}
    \includegraphics[width=0.75\textwidth]{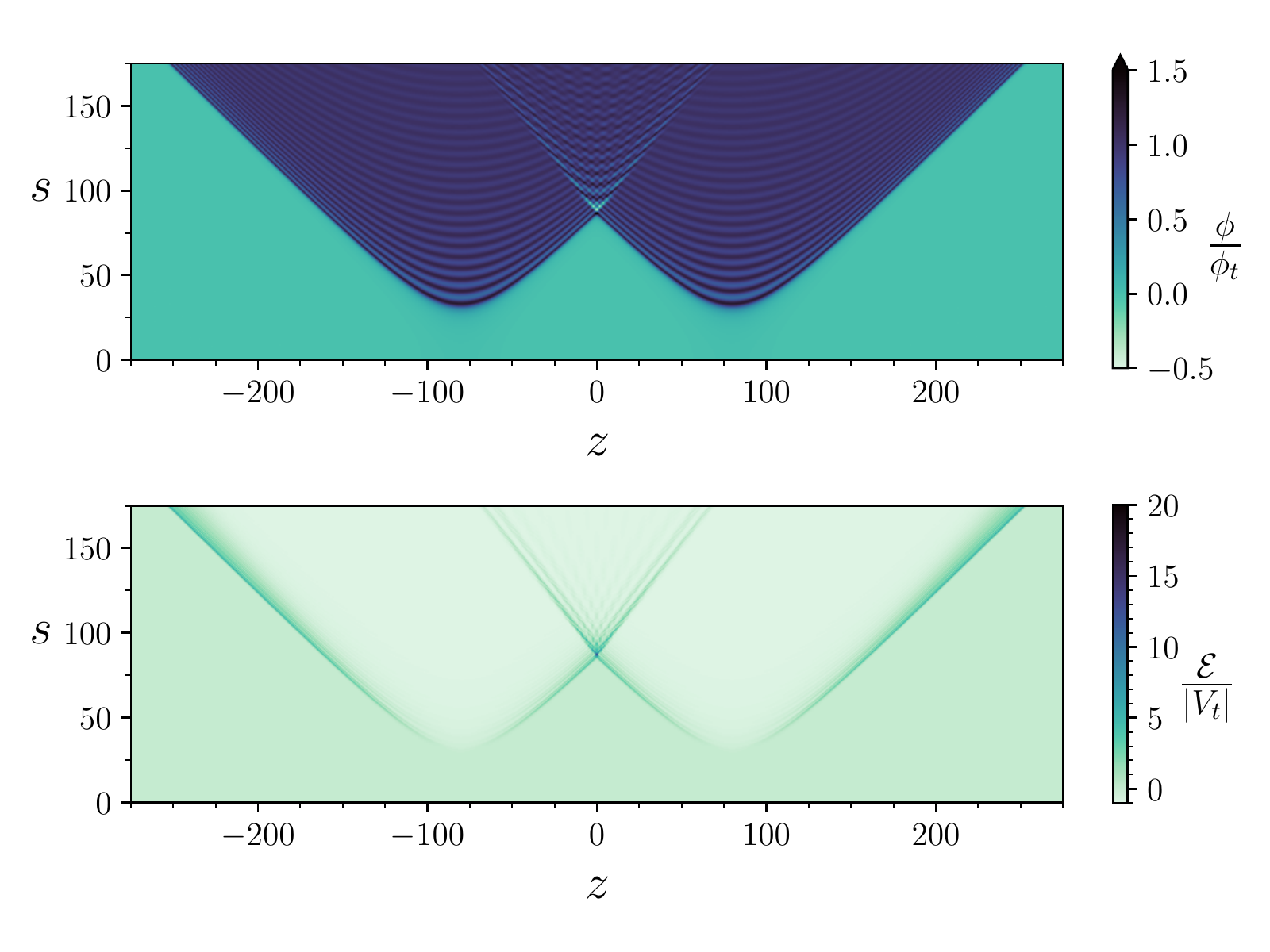}
    \caption{
  Thick-wall bubble collision, $(\lb,\gamma)=(0.01,4)$. }
    \label{fig:field_thick}
  \end{subfigure}
  \caption{The field $\phi$ and energy density $\mathcal{E}$ as a
    function of coordinates $s$ and $z$ for moderate $\gamma$ in the
    thin- and thick-wall regimes. The field and energy density
    have been normalised by their values in the true vacuum.
  \label{fig:field}}
\end{figure*}

We follow Refs.~\cite{Kosowsky:1991ua,Cutting:2018tjt} in normalising the spectrum
\begin{align}
\frac{d\Omega_{\rm GW}}{d\log(\omega)} &\to \frac{1}{(H_* R_* \Omega_{\rm vac})^2}\frac{d\Omega_{\rm GW}}{d\log(\omega)}, \nonumber \\
&=
\frac{1}{\left(\frac{8\pi}{3} d^2\right) \left(\frac{4\pi}{3}s_{\rm max}^3\right) V(\phi_b)^2} \frac{dE_{\rm GW}}{d\log(\omega)},
\label{eq:gw_norm}
\end{align}
to factor out expected scalings.
The quoted values for the fit parameter $\tilde{\Omega}_{\rm GW}$ apply to the scaled spectrum.

\section{Results} \label{sec:results}

In this section, we present the results of our classical simulations of the collisions of two vacuum bubbles, and of their gravitational wave signals.
The parameters for the simulations performed are collected in Appendix~\ref{appendix:table}.
Building on and extending previous studies, we focus on how the dynamics of these collisions depend on two key parameters: $\lb$, which determines the bubble wall thickness (or degree of supercooling), and $\gamma$, the Lorentz factor at collision.
Both $\lb$ and $1/\gamma$ lie in the range $(0,1)$.
We will be particularly interested in $\gamma\gg 1$, which is expected to be relevant to those very strong transitions which yield the largest gravitational wave amplitude.

\subsection{Bubble dynamics} \label{sec:bubble_dynamics_results}

Early studies of bubble nucleation~\cite{Gibbs1876,becker1935kinetische,zeldovich1942theory} were premised upon the thin-wall approximation, which underlies much of our intuition about bubble nucleation and dynamics (see also Ref.~\cite{vehkamaki2006classical}, which uses the thin-wall approximation within classical thermodynamics).
A constant outward pressure, due to the difference in potential energy density between the two phases, causes supercritical bubbles to grow and accelerate, until eventually they collide.

Within this picture, the dynamics of the full field reduces to that of a thin surface, separating regions with different energy density.

Mathematically, the field equations in the thin-wall limit reduce to equations describing the time evolution of the position of the bubble wall, i.e. from partial to ordinary differential equations.
These equations have been formulated and studied in Refs.~\cite{Hawking:1982ga,Wu:1984eda}, and are analytically tractable.
For relativistic two-bubble collisions, the following picture emerges:
The pressure difference between the two phases accelerates the bubble walls until they collide.
At the point of collision, the only way to conserve energy is for the bubble walls to pass through each other, creating a trapped region of the false vacuum

between them.
However, now the pressure is reversed and acts to decelerate the bubble walls, causing them eventually to stop, turn around and then recollide.
This process takes a time
\begin{equation} \label{eq:s_trap}
\frac{s_{\rm trap}}{d} = \left(2^{1/3}-1\right)\sqrt{1-\frac{1}{\gamma^2}},
\end{equation}
and the trapped region is of a spatial extent
\begin{align} \label{eq:z_trap}
\frac{z_{\rm trap}}{d} 
&= 2^{1/3}-1 - \frac{1}{3\gamma} + O\left(\frac{\log(\gamma)}{\gamma^2}\right).
\end{align}
After recollision, the process repeats, with the size of successive trapped regions decreasing.
After many consecutive collisions, the bubble walls eventually become nonrelativistic and cease to recollide.

\begin{figure*}
  \centering
  \setlength{\hfuzz}{1.1\columnwidth}
  \begin{subfigure}{0.45\textwidth}
    \includegraphics[width=0.95\textwidth]{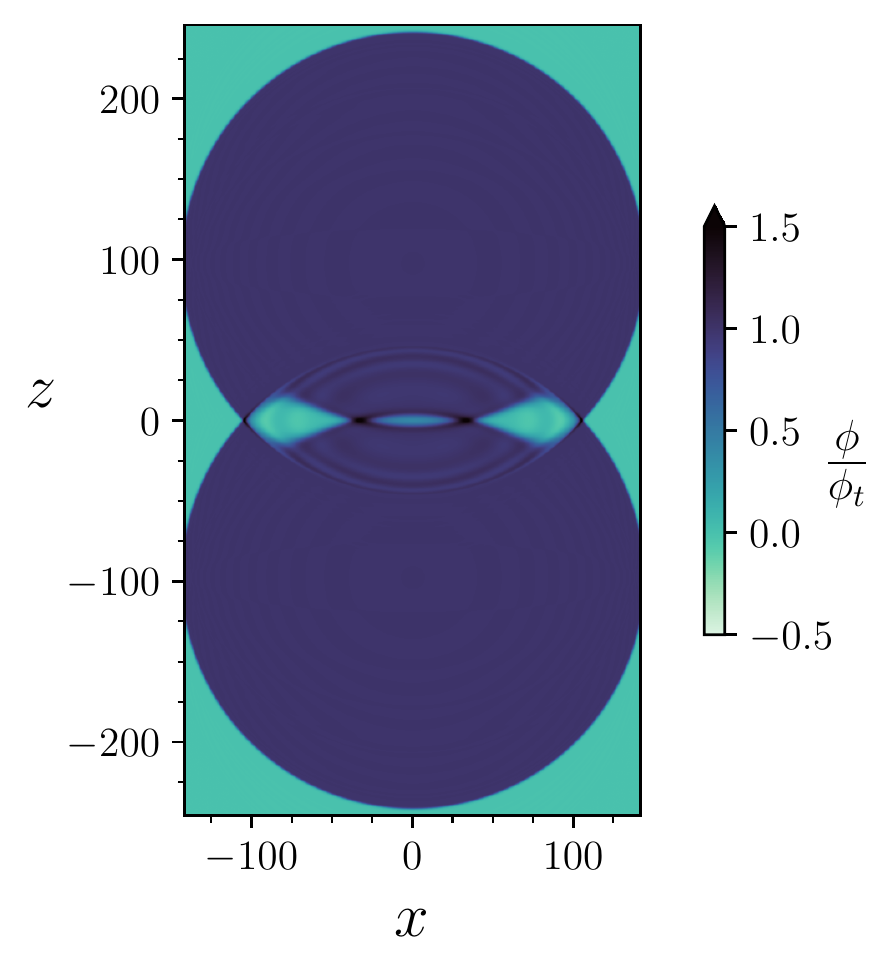}
    \caption{
      $(\lb,\gamma)=(0.9,4)$.
      \label{fig:cylindrical_thin}
    }
    \end{subfigure}
    \begin{subfigure}{0.45\textwidth}
    \includegraphics[width=0.95\textwidth]{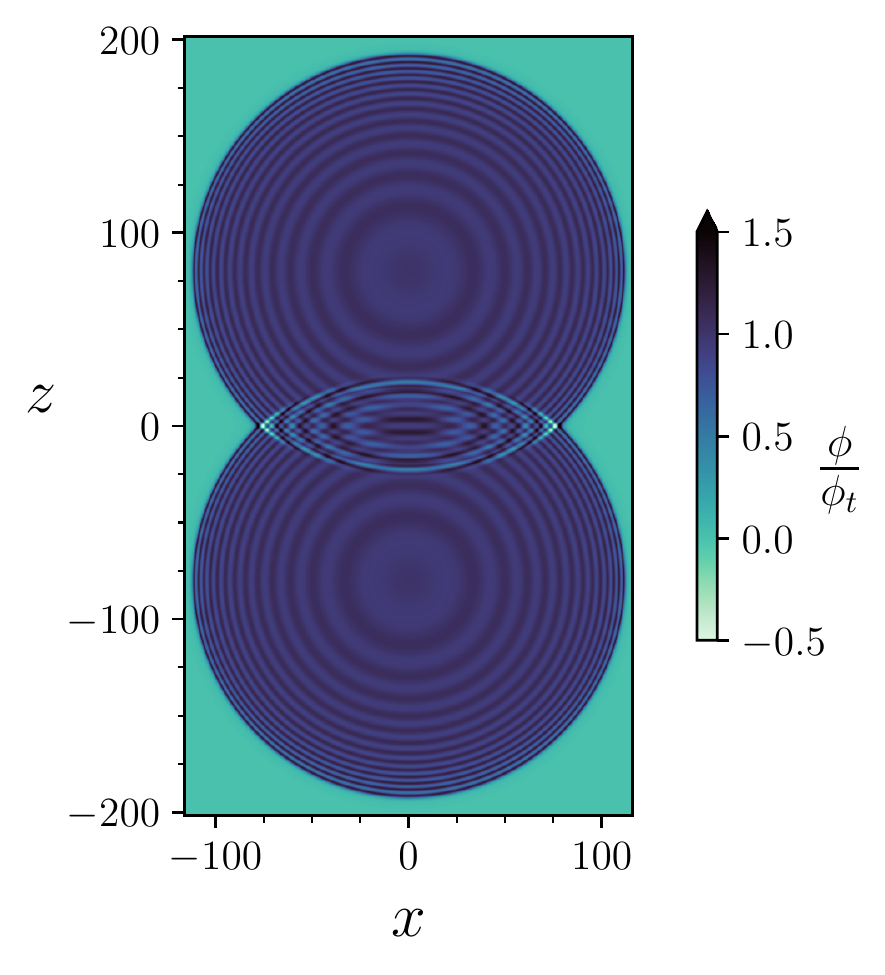}
    \caption{
      $(\lb,\gamma)=(0.01,4)$.
      \label{fig:cylindrical_thick}
    }
    \end{subfigure}
    \caption{
    The bubble collision scenarios plotted in
      Fig.~\ref{fig:field}, shown at time $t = 1.5\ s_\text{col}$
      (see Eq.~\eqref{eq:s_col})
      in cylindrical coordinates; the choice of slicing
      plane $x$ is arbitrary.}
\end{figure*}

As discussed in Sec.~\ref{sec:bubble_dynamics}, parametrically the
thin-wall limit corresponds to $\lb\to 1_-$ in this real scalar
theory.  Fig.~\ref{fig:field_thin} shows the collision of two bubbles
with $(\lb,\gamma)=(0.9,4)$ and reproduces the known thin-wall
behaviour, seen also in Fig.~\ref{fig:cylindrical_thin}.  The trapping
phenomenon is shown clearly in the plot of the field: in the collision
region, approximately diamond-shaped regions of the false vacuum are
produced, as the bubble walls pass through each other before slowing
and bouncing back.  Each successive trapped region is smaller than the
last, and in fact we have verified that Eq.~\eqref{eq:s_trap} holds
rather well.
The lower plot in Fig.~\ref{fig:field_thin}
shows that the energy density is heavily concentrated in the bubble walls.
In addition, one can see that the bubble walls lose energy by
radiating wavelike fluctuations, a phenomenon not captured by the thin
wall limit.

\begin{figure*}[t]
  \centering
  \begin{subfigure}{0.48\textwidth}
    \includegraphics[width=\textwidth]{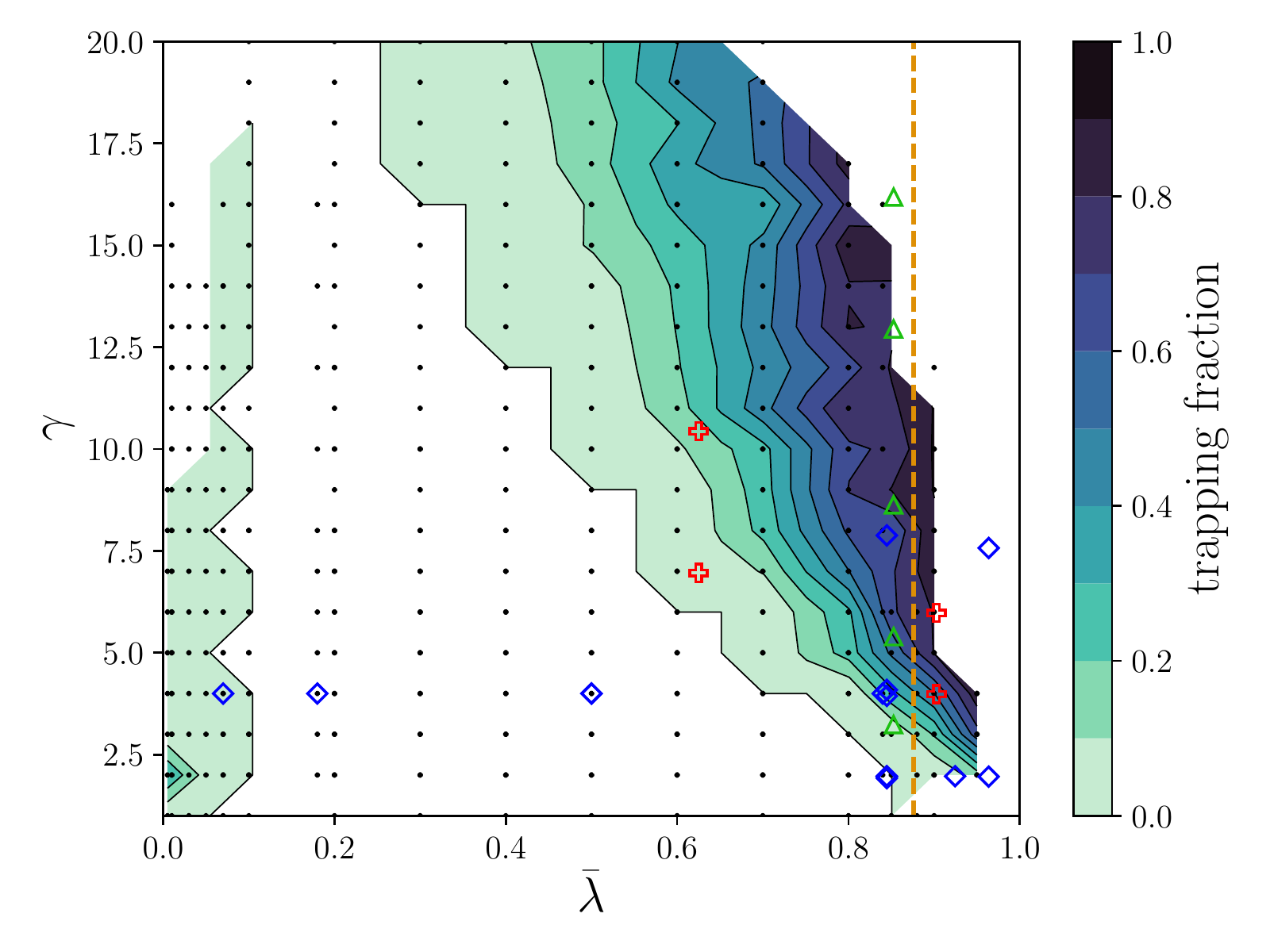}
    \caption{
    }
    \label{fig:trapping}
  \end{subfigure}
  \hfill
  \begin{subfigure}{0.48\textwidth}
  	\vspace{3.4mm}
    \includegraphics[width=\textwidth]{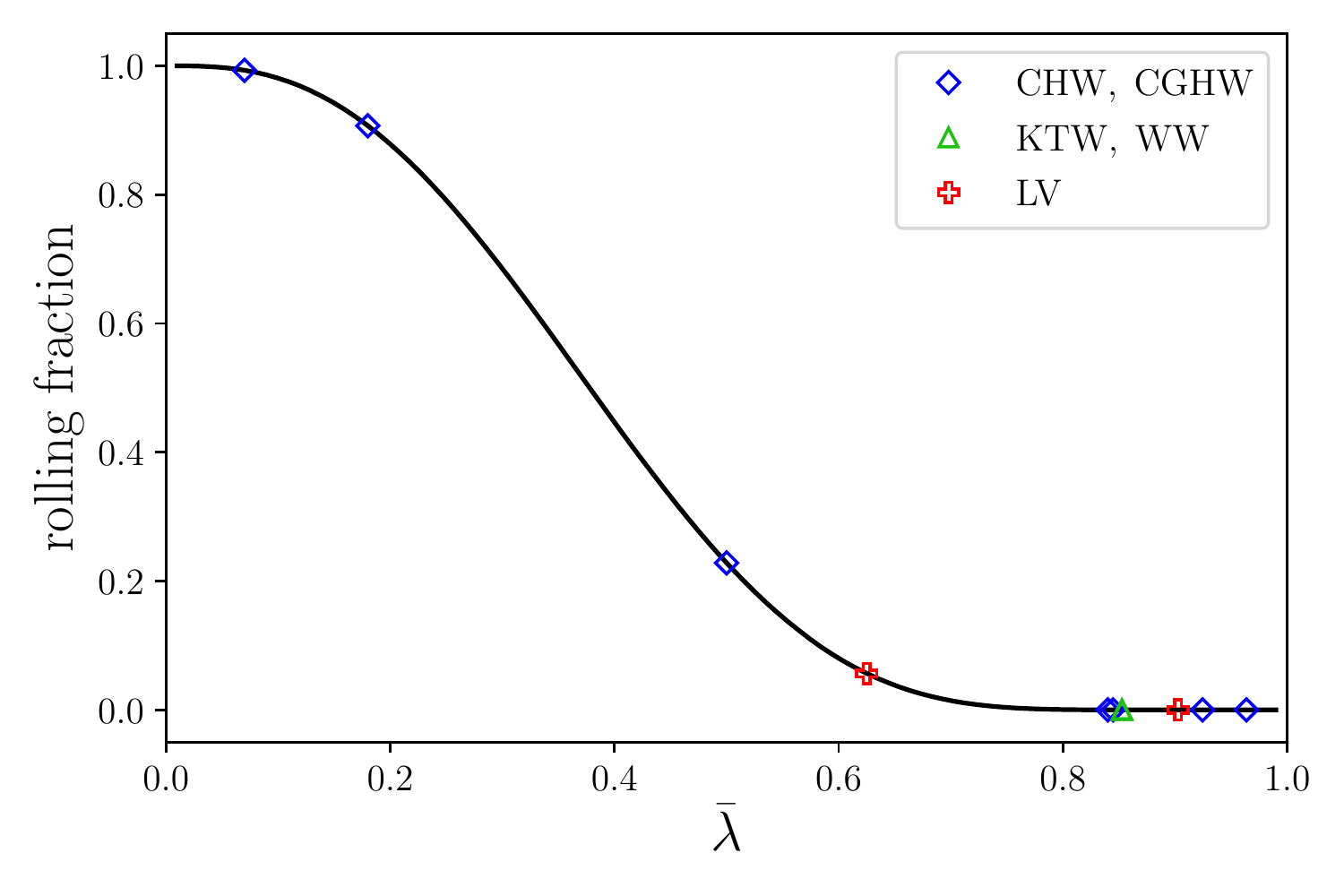}
    \vspace{-0.8mm}
    \caption{
      }
    \label{fig:rolling}
  \end{subfigure}
  \caption{
  Plots of the trapping fraction~(a) and rolling fraction~(b), defined in Eqs.~\eqref{eq:trapping} and \eqref{eq:rolling} respectively.
  Trapping occurs most for ultrarelativistic thin-wall bubbles,
  however also very thick-wall bubbles have a small nonzero trapping fraction.
  In Fig.~\ref{fig:trapping}, data points are shown as black dots, 
  which has been linearly interpolated onto a $20^2$ uniform grid before constructing the contours.  
  The trapping fraction is zero in the central white region.
The dashed orange line shows $\lb=\lbtrap$, to the right of which the one-dimensional {\em trapping equation} of Ref.~\cite{Jinno:2019bxw} predicts trapping to occur.
  Only thick-wall bubbles have a significant rolling fraction.
  In both plots, we also show parameter points studied in the literature:
  	  blue squares from Refs.~\cite{Cutting:2018tjt,Cutting:2020nla},
      green triangles from Refs.~\cite{Kosowsky:1991ua,Watkins:1991zt} and
      red crosses from Ref.~\cite{Lewicki:2019gmv}.
  }
  \label{fig:field_and_energy}
\end{figure*}

Away from the thin-wall limit, trapping occurs less and less.
To quantify this, in our simulations we define the {\em trapping fraction} as the fraction of time that $\phi(s,0)$ spends in the false vacuum after the collision and before the end of the simulation, or mathematically
\begin{align}
\text{trapping fraction} = & \nonumber \\
 \frac{1}{s_{\rm max}-\tilde{s}_{\rm col}}&\int_{\tilde{s}_{\rm col}}^{s_{\rm max}} \theta\left(\phi_{\rm max}-\phi(s,0)\right)\mr{d}s ,
 \label{eq:trapping}
\end{align}
where $\theta$ is the step function, $\phi_{\rm max}$ is the position
of the maximum between phases, and $\tilde{s}_{\rm col}\approx s_{\rm
  col}$ is taken to be the first local maximum in $\phi(s,0)$ after
$s_{\rm col}$
(see Eq.~\eqref{eq:s_col}).
This is plotted in Fig.~\ref{fig:trapping}.
The largest trapping fractions occur for ultrarelativistic thin-wall bubbles,
however very thick-wall bubbles also briefly bounce back to the false vacuum,
with a trapping fraction $\lesssim 0.1$.
Note that the one-dimensional {\em trapping equation} of Ref.~\cite{Jinno:2019bxw} predicts trapping to occur for $\lb\geq \lbtrap$, shown as the dashed orange line in Fig.~\ref{fig:trapping}.%
\footnote{
The definition of trapping from Ref.~\cite{Jinno:2019bxw} corresponds to the limit $s_{\rm max}\to\infty$ of Eq.~\eqref{eq:trapping}, i.e.\ to the infinite time limit. However, while for planar domain walls, trapping may occur for infinite times, for spherical bubbles this does not seem to be the case.
}

\begin{figure*}[t]
\begin{subfigure}{.5\textwidth}
  \includegraphics[width=\textwidth]{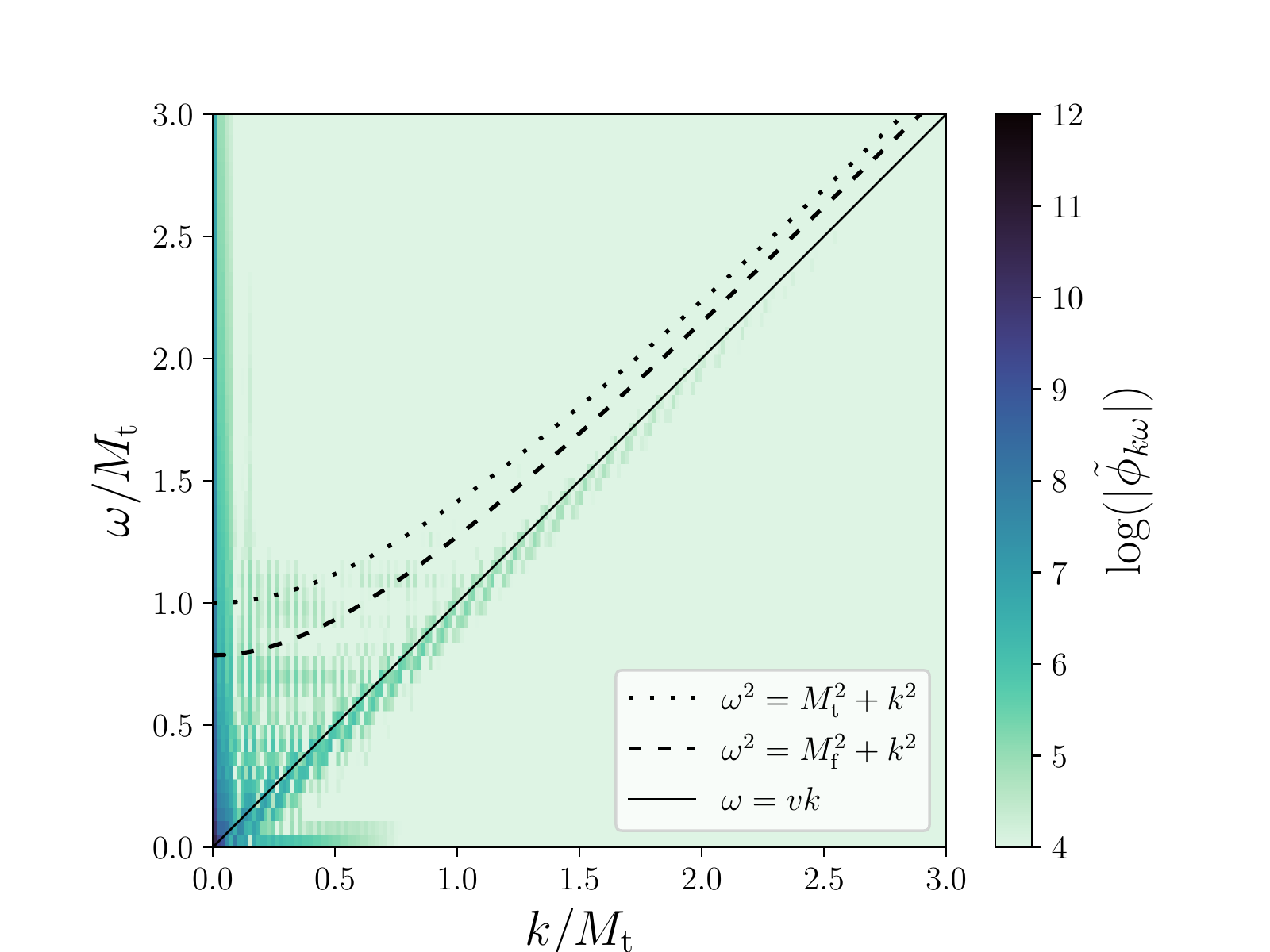}
  \caption{$(\lb,\gamma)=(0.9,4)$}
  \label{fig:fourier_lb_0.9}
\end{subfigure}%
\hfill
\begin{subfigure}{.5\textwidth}
  \includegraphics[width=\textwidth]{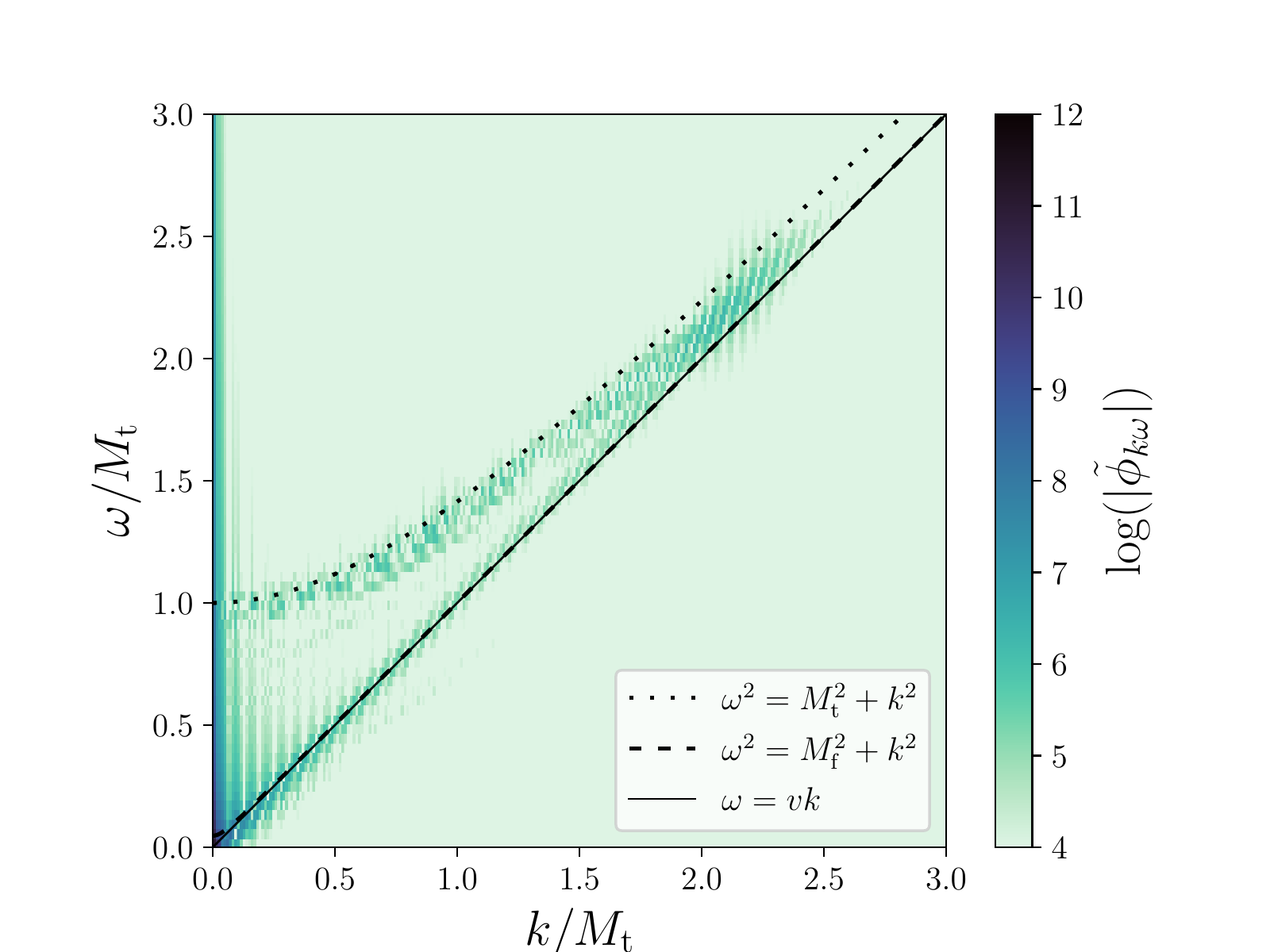}
  \caption{$(\lb,\gamma)=(0.01,4)$}
  \label{fig:fourier_lb_0.01}
\end{subfigure}
  \caption{
  Two-dimensional linear mode expansions of the field $\tilde{\phi}_{k\omega}$ for thick- and thin-wall bubbles.
  In each case only times after the collision of the two bubbles, in the range $s\in [d/2, d]$, are included in the mode expansion.
  Linear excitations of the field about the minima necessarily lie along the dotted and dashed lines shown, while nonlinear excitations need not.
  Note that the linear excitations are significantly more prominent in the thick-wall case.
  The modes along $\omega = v k$ arise because the bubbles are highly relativistic at collision ($v\approx 0.97$), and largely pass through each other.
  The lower left corners contain the low-frequency modes which produce the peak of the gravitational wave spectrum.
  }
  \label{fig:fourier}
\end{figure*}

In the opposite limit, that of thick bubble walls, the dynamics of bubble collisions is qualitatively different.
As one considers thicker and thicker wall bubbles, i.e.\ as $\lb\to 0_+$, the maximum of the potential separating the two minima moves closer and closer to the false vacuum.
The maximum also becomes smaller and smaller relative to the depth of
the true vacuum (as seen in Fig.~\ref{fig:potential}).
The initial bounce configuration in this limit is a roughly Gaussian blob with a very small amplitude (proportional to $\lb$),
just sufficient to peak out beyond the maximum separating the phases.
At nucleation the field is therefore near the top of a tall hill of potential energy, and upon time evolution it rolls down towards the true vacuum.
To quantify this, we define the quantity {\em rolling fraction} as the fraction of the height of the potential energy hill that the field rolls down, or mathematically
\begin{equation} \label{eq:rolling}
\text{rolling fraction} = \frac{V(\phi_0)-V(\phib)}{V(\phi_{\rm max})-V(\phib)},
\end{equation}
where $\phi_0$ is the central value of the bounce configuration.
The rolling fraction is plotted in Fig.~\ref{fig:rolling}, with
points studied in the literature identified.
This reveals that thick-wall bubbles, with a sizeable rolling fraction, have been relatively little studied.

Fig.~\ref{fig:field_thick} shows the collision of two thick-wall bubbles with $(\lb,\gamma)=(0.01,4)$.
Differences from the thin-wall case are immediately apparent in the overall shape of the field and energy density.
At $s=0$, the central value of a thick-wall bubble is far from the true vacuum,
and there is little energy density in the initial condition.
The energy density grows as the field value rolls down the potential energy slope towards the true vacuum, and as it does so, oscillations develop on top of the growing bubbles, forming a wave train in the wake of the bubble wall.
These are visible as the ribbed pattern in the plot of the field in Fig.~\ref{fig:field_thick}.
As the bubble grows and accelerates, these oscillations become more and more Lorentz contracted.

For thick-wall bubble collisions, first the bubble walls collide; then the oscillations in their wakes collide one-by-one.
A significant amount of energy is stored in these oscillations.
This energy density largely passes through the collision centre, approximately along the lightcone, though it appears slowed by the collision.
Upon closer inspection, it can be seen that each oscillation continues
on at close to its collision speed, yet its amplitude damps, thereby
creating the illusion of slowing in Fig.~\ref{fig:field_thick}. 
The first oscillations to collide are also the first to die out after the collision.
Altogether a complicated diffraction-like wave pattern is created
within the future lightcone of the collision centre. This
effect is also clearly seen in Fig.~\ref{fig:cylindrical_thick}.
Unlike for thin-wall bubbles, trapping is all but absent,
as the high-amplitude oscillations in the colliding wave trains
prevent the field from remaining near the false vacuum for long.
Very little energy density remains near the $z=0$ axis after the collision.

To gain additional insight into the difference between the thin- and thick-wall bubble collisions, in Fig.~\ref{fig:fourier} we show the expansion of the field into linear wave modes in each case; see Sec.~\ref{sec:linear_modes}.
The thin-wall case, Fig.~\ref{fig:fourier_lb_0.9}, shows the largest
occupation of modes for small wavenumbers $k \ll \Mb, \Ms$ and small
frequencies $\omega \ll \Mb, \Ms$, in the bottom left corner of the plot.
These modes reflect the structure of $\phi$ at long scales and times, and, as we will see, contribute to the peak of the gravitational wave spectrum.
In addition there is significant occupation of modes along $\omega \approx k$.
This reflects the relativistic bubble walls which pass through each other, moving at an approximately constant speed.
The thick-wall case, Fig.~\ref{fig:fourier_lb_0.01} also shows the largest occupation of modes for small wavenumbers and frequencies.
However, there appear to be fewer structures present in this region than in the thin-wall case.
In addition to this, and in contrast to the thin-wall case, there is a significant occupation of modes along $\omega^2=\Mb^2+k^2$, reflecting the presence of packets of linear excitations about the true vacuum.

A possible dynamical feature that we have not fully explored is the presence or absence of oscillons~\cite{Gleiser:1993pt,Copeland:1995fq}: long-lived, localised nonlinear structures in the scalar field.
Their existence, abundance and longevity depend on the form of the microphysical potential, and they in turn may contribute to the production of gravitational waves~\cite{Amin:2018xfe,Hiramatsu:2020obh}.
However, a static oscillon produced at $z=0$ would break the two boost symmetries of $\mathrm{O(2,1)}$, essentially because it is of an approximately fixed size, and not growing continuously.
This suggests that oscillon production requires collisions of more than two bubbles.
Though we have not found any conclusive evidence of the presence of oscillons, in principle they may be discernible in the Fourier mode decomposition as states lying at just under the $M^2+k^2$ dispersion relation~\cite{Dashen:1975hd,Hindmarsh:2006ur,Zhang:2020bec}.

\subsection{Gravitational waves} \label{sec:gravitational_waves_results}

\begin{figure*}[ht]
\begin{subfigure}{.5\textwidth}
  \includegraphics[width=\textwidth]{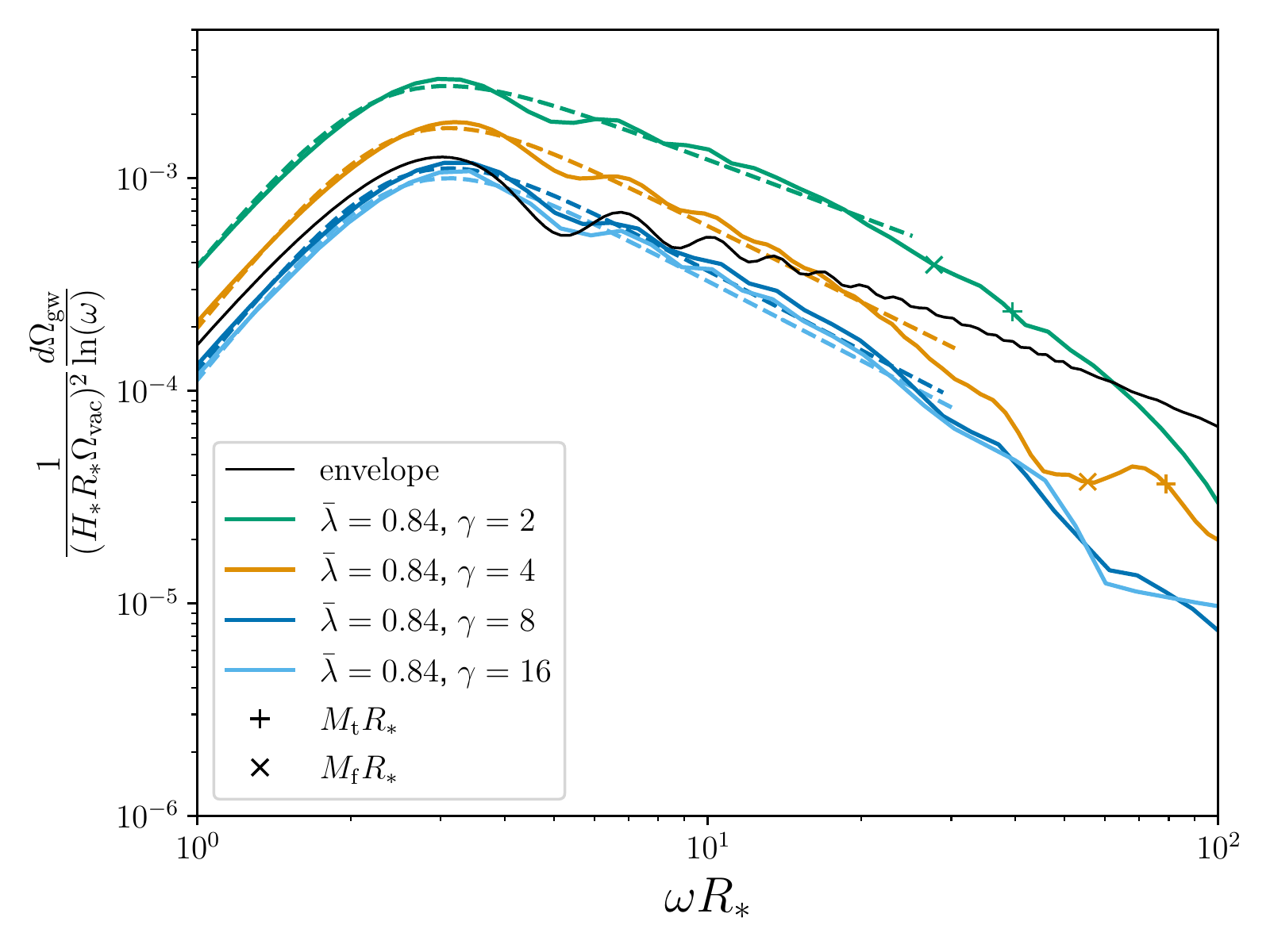}
  \caption{Thin-wall bubbles: $\lb=0.84$.}
  \label{fig:gw_thin}
\end{subfigure}%
\hfill
\begin{subfigure}{.5\textwidth}
  \includegraphics[width=\textwidth]{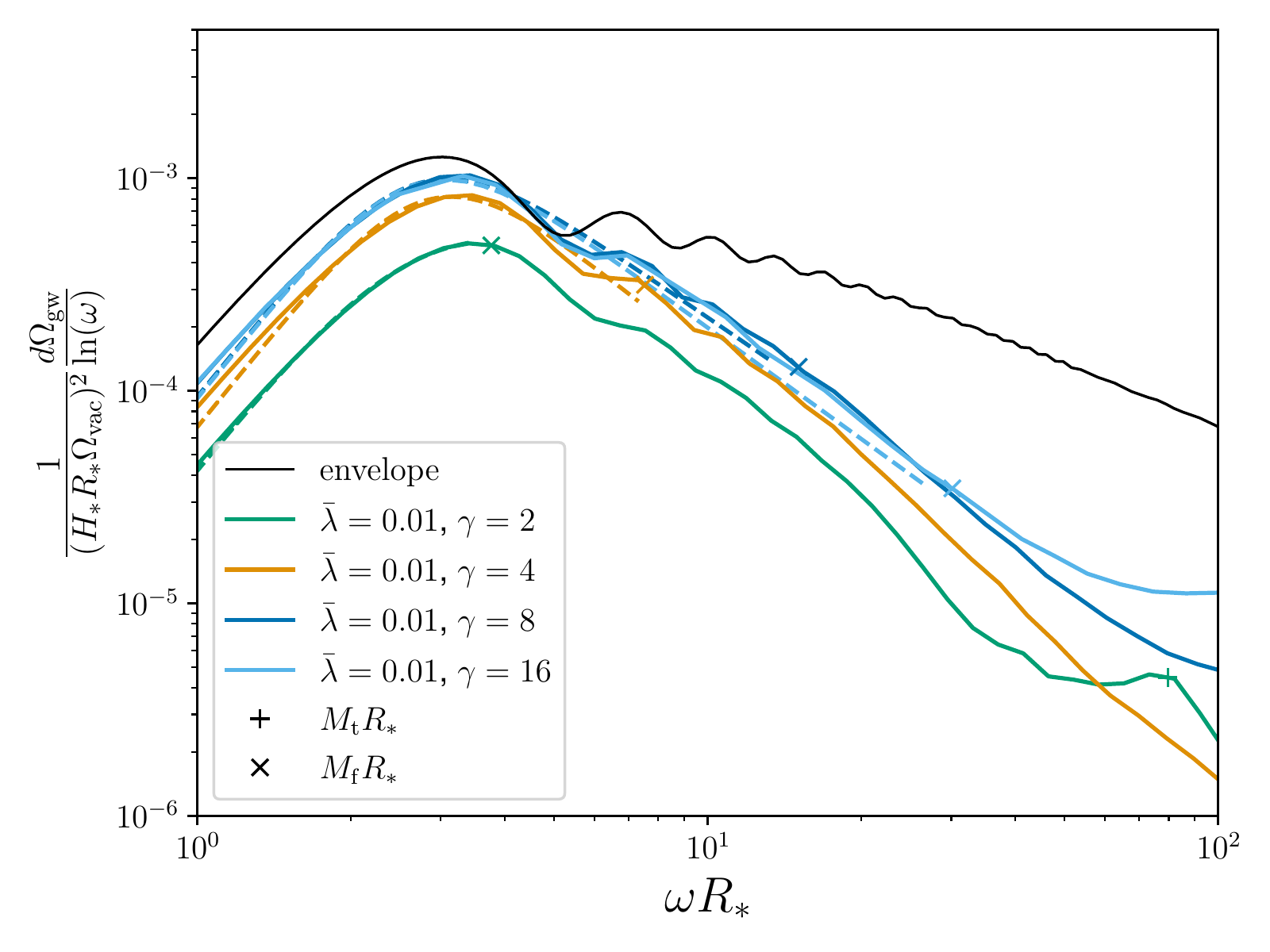}
  \caption{Thick-wall bubbles: $\lb=0.01$.}
  \label{fig:gw_thick}
\end{subfigure}
  \caption{
  Gravitational wave spectra of two-bubble collisions.
  Fits (dashed) using Eq.~\eqref{eq:fit_fn} are shown alongside the numerical data (full), as well as the result of the envelope approximation (black).
  The plusses and crosses show the location of the masses in the true and false vacua respectively.
  Note that the uptick visible in the $(\lb,\gamma)=(0.01,16)$ line for $\omega R_* \gtrsim 50$ is a lattice artefact: decreasing the lattice spacing moves this feature to larger values of $\omega R_*$.
  }
  \label{fig:gw_examples}
\end{figure*}

The dynamics of thin-wall and thick-wall bubble collisions are rather different, as we have demonstrated above.
This difference is determined by the parameters of the theory on microphysical scales, yet it may be observable today on macroscopic scales if it has a significant effect on the resulting GW signals.
In this section, we present our results for the GW spectra of two-bubble collisions, using the formalism outlined in Sec.~\ref{sec:gravitational_waves}.
We focus our attention on how the spectra depend on the parameters $\lb$ and $\gamma$.

For each studied parameter point in the $(\lb,\gamma)$ plane, we have calculated the GW spectrum at a number of angular frequencies, typically 61, evenly spaced in log-space in the range $[\omega_{\rm min},\omega_{\rm max}]$. 
We used $\omega_{\rm min}=\pi/L_z$, where $L_z$ is the size of the simulation lattice in the $z$-direction, and $\omega_{\rm max}=\text{min}(\pi/\delta z,10 \Mb)$.

Fig.~\ref{fig:gw_examples} shows the GW spectra calculated for two values of $\lb$, one thin-wall with $\lb=0.84$ and one thick-wall with $\lb=0.01$.
In common with the literature on GWs, we normalise the frequency with $R_*$,
which for two-bubble collisions we may identify with the input parameter $d$.
Lorentz factors $\gamma=2,4,8,16$ are plotted together.  For
comparison, the GW spectrum from the envelope approximation is shown in
black~\cite{weir_david_2021_5090669}.

For a fixed value of $\lb$, it can be seen that the spectra appear to
converge as the Lorentz factor grows.  At large enough Lorentz
factors, the dependence on the Lorentz factor is accounted for by the
overall scalings of Ref.~\cite{Kosowsky:1991ua}: the peak frequency
scales as $\omega_{\rm peak}\propto \gamma^{-1}$ and the peak
amplitude as $\Omega_{\rm peak}\propto \gamma^5$.  Further, the values
of the peak frequency and amplitude agree relatively well with
the prediction of the envelope approximation.

There are clear differences between the thin- and thick-wall spectra in Figs.~\ref{fig:gw_thin} and \ref{fig:gw_thick}.
For the smaller Lorentz factors studied, both the amplitude and the high-frequency slope $\omega^{-b}$ of the spectra differ significantly.
At large Lorentz factors, the spectra in Figs.~\ref{fig:gw_thin} and \ref{fig:gw_thick} appear to converge towards a similar peak amplitude.
However, the high-frequency slope of the thick-wall spectrum is steeper even at large Lorentz factors.

\begin{figure*}[t]
  \centering
  \begin{subfigure}{0.48\textwidth}
  \includegraphics[width=\textwidth]{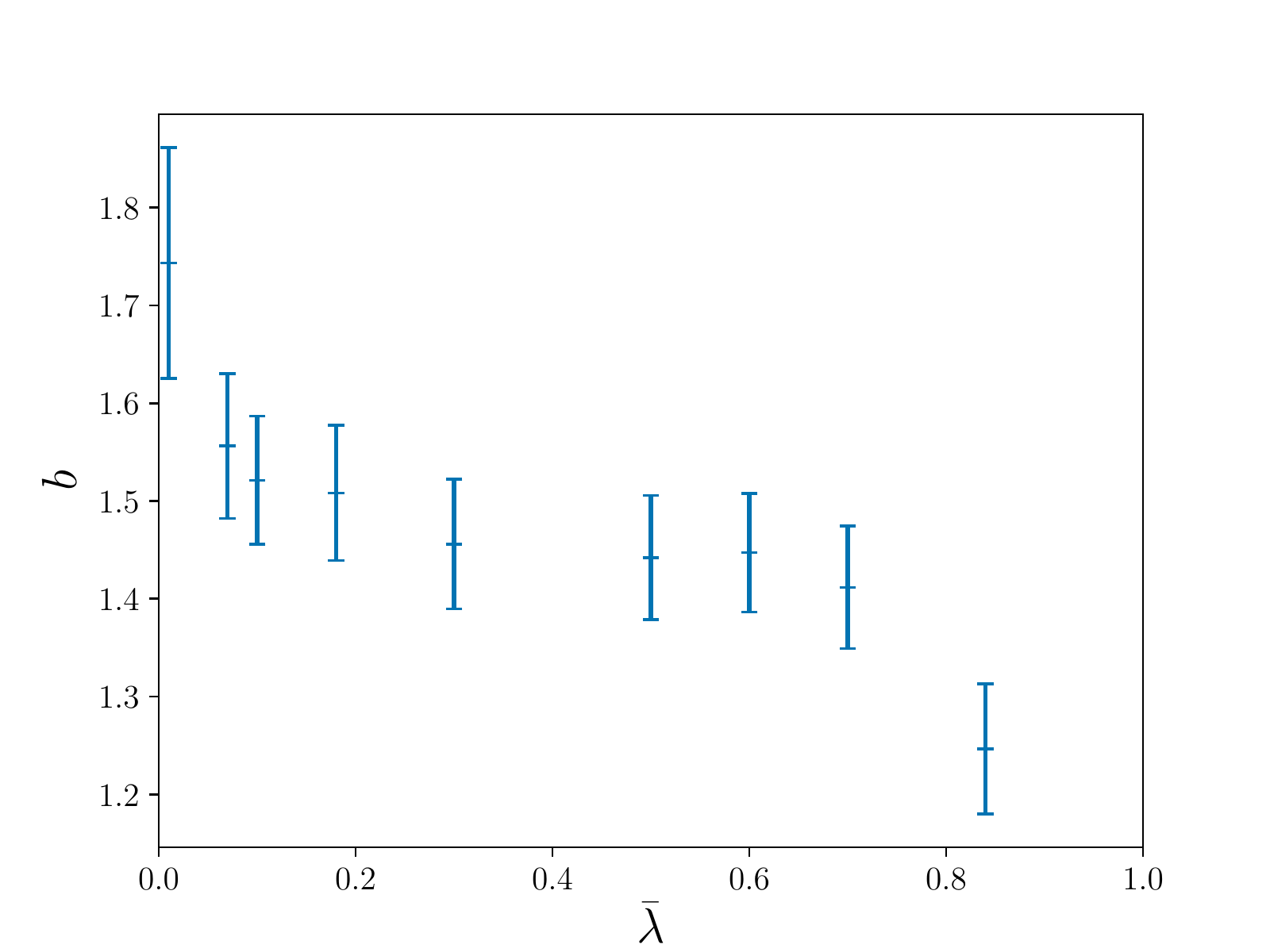}
  \caption{$\gamma=16$}
  \label{fig:b_16}
  \end{subfigure}
  \hfill
  \begin{subfigure}{0.48\textwidth}
  \includegraphics[width=\textwidth]{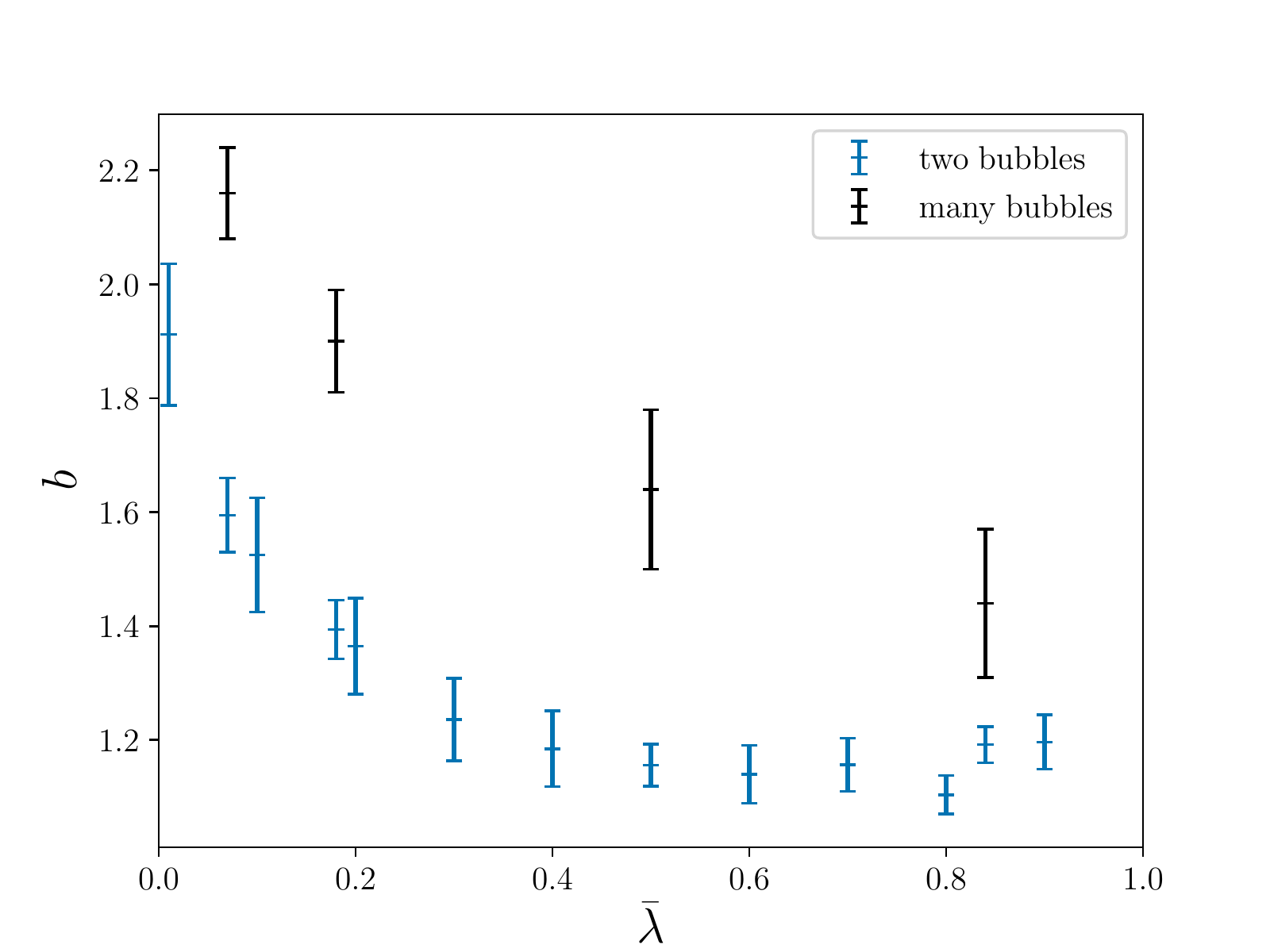}
  \caption{$\gamma=4$}
  \label{fig:b_4}
  \end{subfigure}
  \caption{
  The fit parameter $b(\lb,\gamma)$ of Eq.~\eqref{eq:fit_fn}, the high-frequency power law of the GW spectrum, here plotted together with the fit errors at two fixed values of the Lorentz factor.
  In Fig.~\ref{fig:b_16} the results for the largest Lorentz factor studied, $\gamma=16$, are plotted against the parameter $\lb$ determining the degree of supercooling.
  In Fig.~\ref{fig:b_4} the results for $\gamma=4$ are plotted together with those from Ref.~\cite{Cutting:2020nla} for many-bubble collisions.
  }
  \label{fig:b_fixed_gamma}
\end{figure*}

\begin{figure*}[t]
  \begin{subfigure}{0.5\textwidth}
  \vspace{-4.8mm}
  \includegraphics[width=\textwidth]{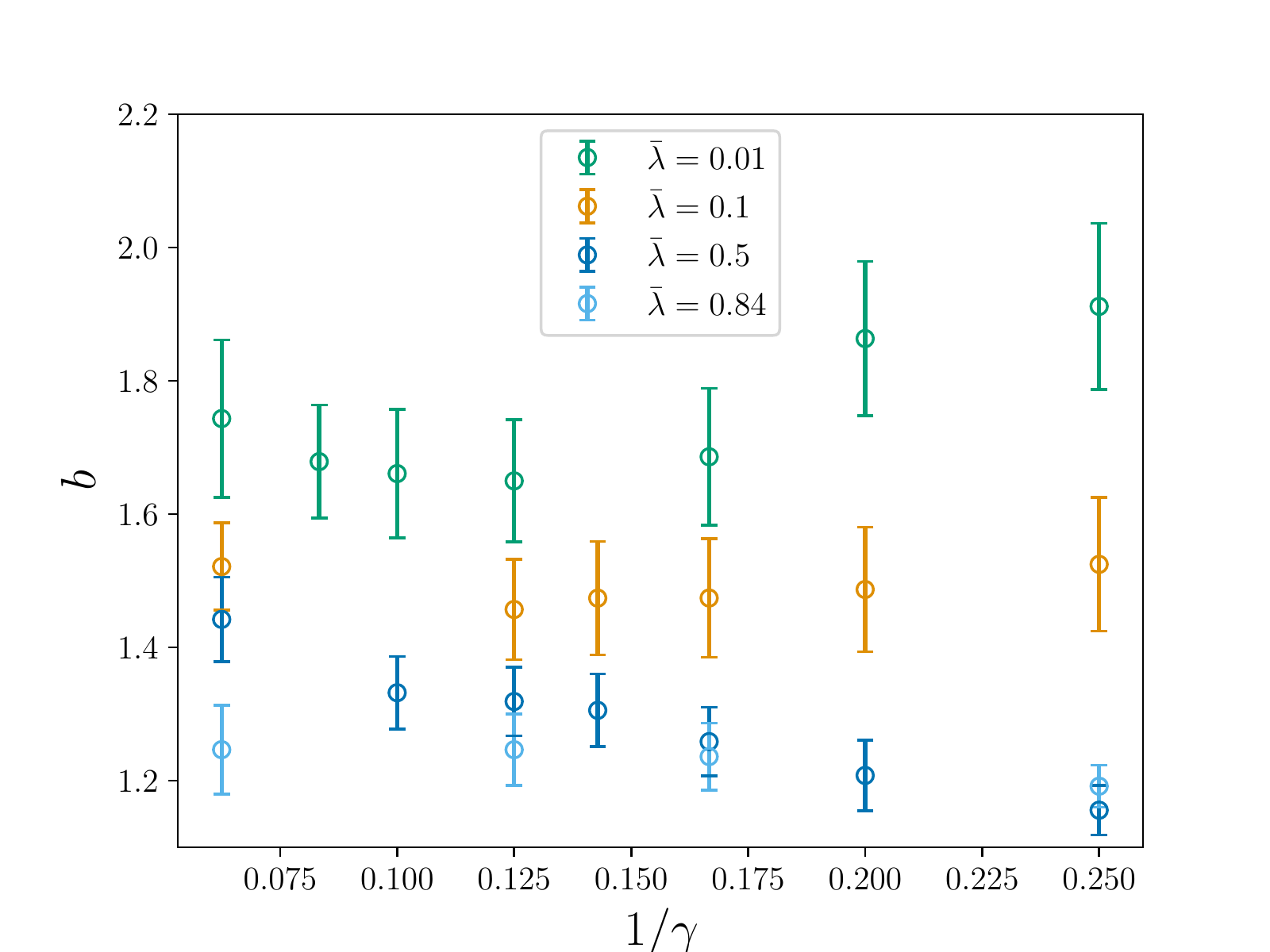}
  \vspace{-2.3mm}
  \caption{}
  \label{fig:b_gamma}
  \end{subfigure}
  \hfill
  \begin{subfigure}{0.4755\textwidth}
  \includegraphics[width=\textwidth]{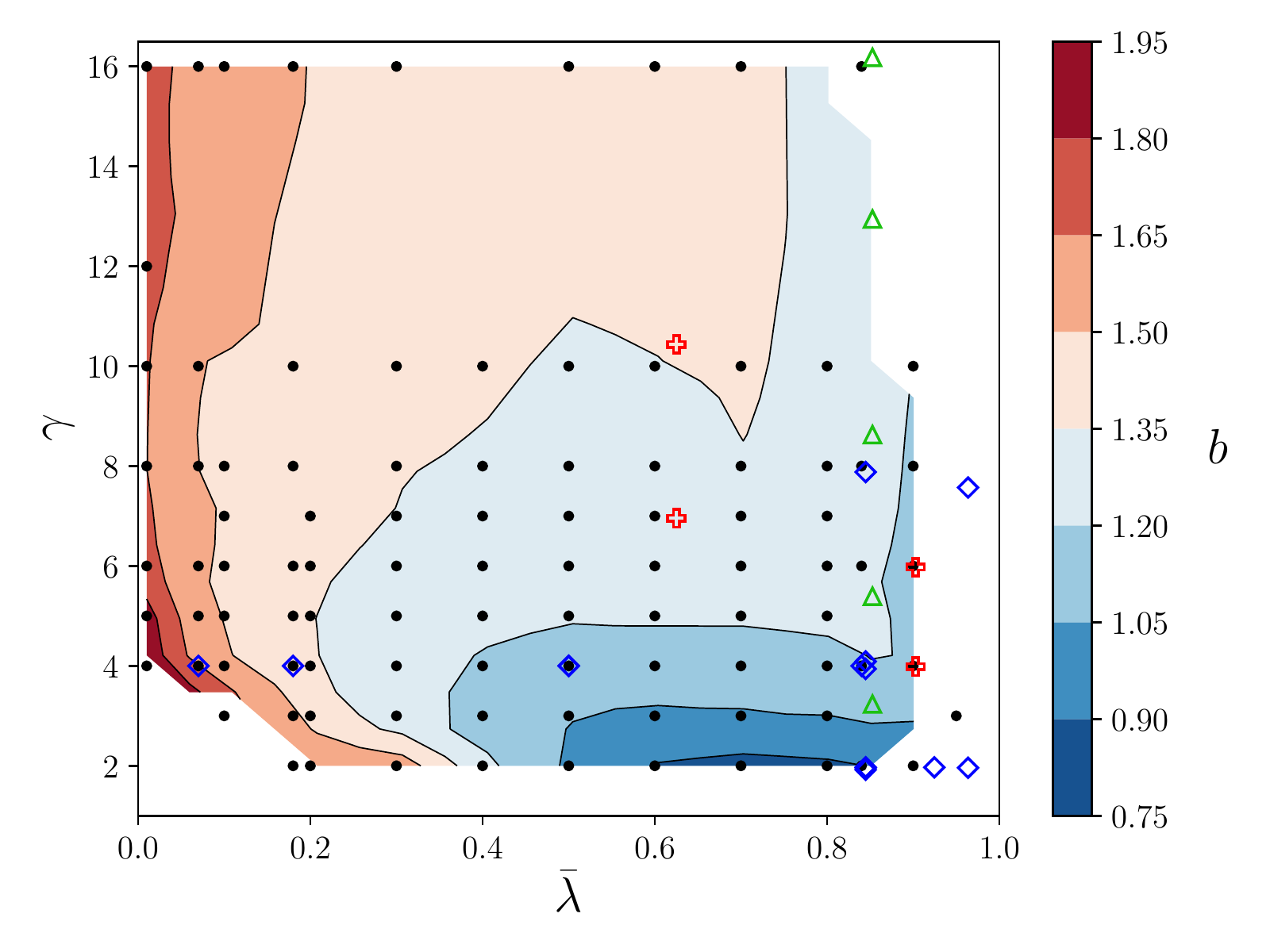}
  \caption{}
  \label{fig:b_contour}
  \end{subfigure}
  \caption{
  The fit parameter $b(\lb,\gamma)$ of Eq.~\eqref{eq:fit_fn}, the high-frequency power law of the GW spectrum.
  In Fig.~\ref{fig:b_gamma} the results are plotted against the inverse Lorentz factor, $1/\gamma$, for four different values of the parameter $\lb$.
   Fig.~\ref{fig:b_contour} summarises our results for $b(\lb,\gamma)$. 
  Black dots denote the locations of our numerical data, which have been linearly interpolated onto a $20^2$ uniform grid before constructing the contours.
  Some data points in the lower left corner have been omitted as the peak
  frequency is too near the fit cut off $\omega_{\rm cut}$.  
  In Fig.~\ref{fig:b_contour} we also show the locations of simulations carried out in the literature:
      blue squares from Refs.~\cite{Cutting:2018tjt,Cutting:2020nla},
      green triangles from Refs.~\cite{Kosowsky:1991ua,Watkins:1991zt} and
      red crosses from Ref.~\cite{Lewicki:2019gmv}.
  }
  \label{fig:b}
\end{figure*}

Figs.~\ref{fig:b_fixed_gamma} and \ref{fig:b} quantify how the exponent $b$ of the high-frequency slope varies with $\lb$ and $\gamma$.
Fig.~\ref{fig:b_contour} summarises these results in a contour plot of $b$ across the $(\lb,\gamma)$ plane.
This reveals a great deal of structure.
The most shallow high-frequency slopes, with $b\approx 0.9$, are produced by relatively slow moving thin-wall bubbles, in the lower right-hand corner of the contour plot.
The steepest high-frequency slopes, with $b\approx 1.9$ are produced by relatively slow moving thick-wall bubbles, in the lower left hand corner of the contour plot.
As the Lorentz factor grows, the differences between thin and thick walls become less pronounced.
However, even at Lorentz factors as large as $\gamma=16$, a significant difference remains.
This can be seen in Figs.~\ref{fig:b_16} and \ref{fig:b_gamma}, together with the estimated fit errors.
Fig.~\ref{fig:b_4} shows how the exponent $b$ for two-bubble collisions compares to that for many-bubble collisions, taken from Ref.~\cite{Cutting:2020nla}.
While the $\lb$-dependence agrees qualitatively between the two cases, the high-frequency slope is somewhat larger for many-bubble collisions.

The other two fit parameters for the gravitational wave spectrum are shown in Appendix~\ref{appendix:extra_fits}.
In both cases there is a significant amount of structure at small Lorentz factors,
which washes out as the Lorentz factor increases.
Notwithstanding, at large Lorentz factors, the peak frequency $\tilde{\omega}$ is marginally higher for thin-wall bubbles, and the peak amplitude $\tilde{\Omega}_{\rm GW}$ is marginally smaller for intermediate thickness bubble walls.
The high-frequency exponent $b$ shows the strongest dependence on $\lb$ at large Lorentz factors.

The results of Sec.~\ref{sec:bubble_dynamics_results} on the scalar field dynamics suggest some possible explanations for the differences in the GW spectra of thin and thick-wall bubble collisions.
In Fig.~\ref{fig:fourier} it was shown that only thick-wall bubbles show a significant occupation of linear modes about the true vacuum, perhaps due to the oscillations initiated through rolling down the potential barrier.
An arbitrary superposition of linear scalar modes, satisfying $\omega^2=M^2+k^2$, does not source GWs at $O(G_N)$, simply due to kinematics, and hence their presence would naturally lead to a reduced gravitational wave amplitude, at least at these larger wavenumbers.
In addition, the phenomenon of trapping, which occurs predominantly for thin-wall bubbles, is a time-dependent nonlinear phenomenon with the potential to source significant GWs at frequencies higher than $\omega_{\rm peak}$.
Each of these factors, or their combination, may explain the steeper high-frequency power law produced by thick-wall bubble collisions.

\section{Conclusions} \label{sec:conclusions}

In this article, we have studied vacuum two-bubble collisions and their GW spectra, focusing on the dependence on the Lorentz factor $\gamma$ and the microphysical Lagrangian parameter $\lb$, which determines how thick or thin the bubble walls are at nucleation.
In agreement with previous studies, we have found that at fixed $\lb$ and as $\gamma\to \infty$ the GW spectrum appears to converge towards a fixed spectrum, up to known scalings.
However, the converse is not true.
At fixed Lorentz factors, even at $\gamma\gg 1$, we have shown that the GW spectrum depends significantly on $\lb$, which determines how thick or thin the bubble walls are at nucleation.
This corroborates the conclusions of Ref.~\cite{Cutting:2020nla} at higher Lorentz factors.
In particular we have shown that the high-frequency power law $\omega^{-b}$ is steeper than that of the envelope approximation, which for two-bubble collisions is $b=0.88\pm0.02$, varying between $b=1.74\pm 0.12$ and $b=1.25\pm 0.07$ as $\lb$ varies from 0.01 to 0.84 at $\gamma=16$;
see Fig.~\ref{fig:b_16}.

This conclusion is perhaps quite surprising, as the GW spectrum peaks at frequencies of order $1/R_*$, much smaller than the frequencies of particle oscillations $M\gg 1/R_*$ which characterise the underlying Lagrangian parameters.
Thus, microphysics and macrophysics do not decouple in vacuum bubble collisions;
the value of $\lb$ determines large-scale qualitative features of the bubble collision dynamics.

We have characterised these large-scale features in a variety of ways.
The phenomenon of trapping occurs for thin-wall bubbles at $\lb \approx 1$.
On the other hand new oscillatory phenomena appear for thick-wall bubbles at $\lb\ll 1$, as a result of the field rolling down the true vacuum potential well, after nucleation.
These phenomena have discernible effects on the overall shape of the field evolution, on its energy density and on its Fourier mode decomposition.

While our simulations were performed only for two-bubble collisions, our qualitative conclusions should hold also in many-bubble collisions, as a result of the presence of the same underlying physical phenomena.
However, the power laws of the GW spectrum will differ for many-bubble collisions.
In fact Ref.~\cite{Cutting:2020nla} found that the high-frequency power law $b$ in many-bubble collisions has an even stronger dependence on $\lb$ than we have found in the two-bubble case, as can be seen in Fig.~\ref{fig:b_4}.
Thus, more work is needed in future to determine these power laws for many-bubble collisions at larger Lorentz factors.

In the full many-bubble simulations of Ref.~\cite{Cutting:2020nla}, long-lived, localised fluctuation regions appear to be present after the bubbles have coalesced; for a video of the simulation see \cite{CuttingVideo}.
We conjecture that field oscillations in these regions are responsible for the formation of a gravitational wave peak at the mass scale in those simulations.
Furthermore, if these regions are nascent oscillons, they are expected to rapidly become spherical~\cite{Kolb:1993hw,Hindmarsh:2006ur} and would then cease to source gravitational waves.
As discussed at the end of Sec.~\ref{sec:bubble_dynamics_results}, these localised regions do not expand with time, and hence do not obey the $\mathrm{O(2,1)}$ symmetry of two-bubble collisions.
Therefore, the timescale on which these processes occur, and their broader importance, are deferred to future work on many-bubble collisions.

In summary, in the stochastic
gravitational wave background of a vacuum first-order phase transition, we have shown that the high-frequency power law $\omega^{-b}$ depends on $\lb$, a microphysical Lagrangian parameter.
This extends the scope of GW experiments to probe particle physics in the early universe, by breaking otherwise limiting degeneracies~\cite{Gowling:2021gcy}.

\section*{Acknowledgements}
The authors would like to thank
Daniel Cutting,
Mark Hindmarsh,
Ryusuke Jinno,
Paul Saffin,
and Essi Vilhonen
for enlightening discussions, and valuable comments on the manuscript.
O.G. (ORCID ID 0000-0002-7815-3379) was supported by the Research Funds of the
University of Helsinki, and U.K.\ Science and Technology Facilities
Council (STFC) Consolidated grant ST/T000732/1.
S.S. (ORCID ID 0000-0003-0475-3395) was supported by
the Research Funds of the University of Helsinki, by the Helsinki
Institute of Physics, and by Academy of Finland grant no. 328958.
D.J.W. (ORCID ID 0000-0001-6986-0517) was supported by Academy of
Finland grant nos. 324882 and 328958.  Simulations for this paper were
carried out at clusters provided by: the Finnish Grid and Cloud
Infrastructure at the University of Helsinki
(urn:nbn:fi:research-infras-2016072533), CSC – IT Center for Science,
Finland, and the University of Nottingham's Augusta HPC service.

\appendix

\section{Potential conventions} \label{appendix:conventions}

For ease of comparison with other works, we list here the relations between the convention of Refs.~\cite{Enqvist:1991xw,Cutting:2020nla}, which we adopt, and some other conventions in the literature.
The relation to the convention of Refs.~\cite{Kosowsky:1991ua,Kosowsky:1992rz,Kosowsky:1992vn,Watkins:1991zt,Child:2012qg} is
\begin{align}
&\lb = \frac{3}{2} + \nonumber \\
& \frac{3}{2}\left[\frac{2
	+ \cos \left(\frac{4\cos^{-1}(\varepsilon)}{3}\right)
	+ \sqrt{3} \sin \left(\frac{4\cos^{-1}(\varepsilon)}{3}\right)}
	{2
	- \cos \left(\frac{2\cos^{-1}(\varepsilon)}{3}\right)
	+\sqrt{3} \sin \left(\frac{2\cos^{-1}(\varepsilon)}{3}\right)}-4\right]^{-1},\\
\varepsilon &= 3\sqrt{3}\epsilon,
\end{align}
where the phase transition occurs for $\epsilon \in (0,1/(3\sqrt{3}))$.
In these references, most simulations were carried out for $\epsilon=0.1 \Rightarrow \lb\approx 0.853$, and hence in the relatively thin-wall regime.
The relation to the convention of Refs.~\cite{Jinno:2019bxw,Lewicki:2019gmv} is,
\begin{equation}
\lb = \frac{a(a+3)}{(a+2)^2},
\end{equation}
where the phase transition occurs for $a \in (0,\infty)$.
The two simulations of Ref.~\cite{Lewicki:2019gmv} were carried out for $a=2,10 \Rightarrow \lb \approx 0.625, 0.903$, in the intermediate and thin-wall regimes respectively.

Ref.~\cite{Bond:2015zfa} carried out two runs using two different potentials, which can be related to $\lb$ via the conventions given above.
Explicitly, for their linear potential $\delta_{\rm linear} = 2\epsilon$ and for their cubic potential $\delta_{\rm cubic} = 3/(3 + a)$.
Their thin- and thick-wall runs were equivalent to $\lb\approx 0.941$ and $\lb\approx 0.0223$ respectively.

\section{Discrete equation of motion}
\label{appendix:discrete_eom}

For the lattice simulation of the scalar field, we discretise the equation of motion, Eq.~\eqref{eq:eom}, in the form
\begin{align}
\phi_{s+1,z} & = 
\phi_{s,z}+
\Pi_{s+\frac{1}{2},z} \delta s, \\
\Pi_{s+\frac{1}{2},z} & =
\left(\frac{s - \delta s}{s+ \delta s} \right) 
\Pi_{s-\frac{1}{2},z}
-\frac{s\ \delta s}{(s + \delta s)}\frac{\partial V}{\partial \phi}(\phi_{s,z})
\nonumber\\
&\quad +\frac{s\ \delta s}{s + \delta s} 
\left(
\frac{\phi_{s,z+1} - 2\phi_{s,z} + \phi_{s,z-1}}{\delta z^2} \right)
,
\end{align}
where $\Pi= \frac{d\phi}{ds}$ denotes the momentum conjugate to $\phi$.

\section{Discrete mode expansion} \label{appendix:fourier}

For the lattice-discretised field, we utilise a discrete mode expansion which reduces to that of Sec.~\ref{sec:linear_modes} in the continuum limit.
Here we take $z$ and $s$ to be integers labelling the lattice sites, and running over the ranges $[0,N_z)$ and $[0,N_s)$ respectively.
In the $z$-direction the transform is a type-I discrete cosine transform, and in the $s$-direction it is a sinc-type transform.
The transform is orthogonal with weight $(s+1)^2$.
Explicitly, it takes the form:
\begin{align}
\tilde{\phi}_{k\omega} &=
\sum_{z=0}^{N_z-1} \sum_{s=0}^{N_s-1}
\sigma_{z} (s + 1)^2 f_{k\omega, z s} \phi_{z s}, \\
\phi_{z s} &=
\sum_{k=0}^{N_z-1} \sum_{\omega=0}^{N_s-1}
\sigma_{z} f_{z s, k\omega} \tilde{\phi}_{k \omega},
\end{align}
where the discrete Fourier modes are
\begin{align}
f_{ia, jb}  &= 
\frac{2}{\sqrt{N_s(N_z-1)}}
\frac{1}{b}
\cos\left(\frac{\pi i j}{N_z-1}\right)
\sin\left(\frac{\pi a b}{N_s}\right),
\end{align}
and we have introduced
\begin{equation}
\sigma_{z} = 1-\frac{1}{2}\delta_{z, 0}-\frac{1}{2}\delta_{z, N_z-1}.
\end{equation}
This mode expansion is utilised in Fig.~\ref{fig:fourier}.

\section{Numerical tests} \label{appendix:tests}

The numerical results of this article rely chiefly on three numerical computer codes~\cite{sukuvaara_satumaaria_2021_5127538}, written in {\tt Python}, which respectively evolve the scalar field, calculate the discrete mode expansion of the scalar field, and perform the integrals for calculating the gravitational wave spectrum.
Here we report the results of consistency and convergence tests performed on these three codes.

A common test for simulation codes performing time evolution is to test the conservation of energy.
However, due to the damping term in Eq.~\eqref{eq:eom}, the evolution
of the scalar field does not conserve `energy' on constant $s$-slices.%
\footnote{
    Here, by `energy', we refer simply to the sum of scalar kinetic and potential energy density terms integrated over a surface of constant $s$.
	This is not conserved because translations in $s$ are not a symmetry.
	Of course, the true energy corresponding to translations in $t$ is conserved.
}
Instead, one can test the rate of decay of the energy, which can be shown to be~\cite{Kosowsky:1991ua}
\begin{equation} \label{eq:energy_derivitive}
\frac{dE}{ds} = -\frac{4\pi}{s}\int_{-\infty}^{\infty}dz \left(\frac{\partial \phi}{\partial s}\right)^2.
\end{equation}

\begin{figure}[t]
\setlength{\hfuzz}{1.1\columnwidth}
  \centering
    \includegraphics[width=0.48\textwidth]{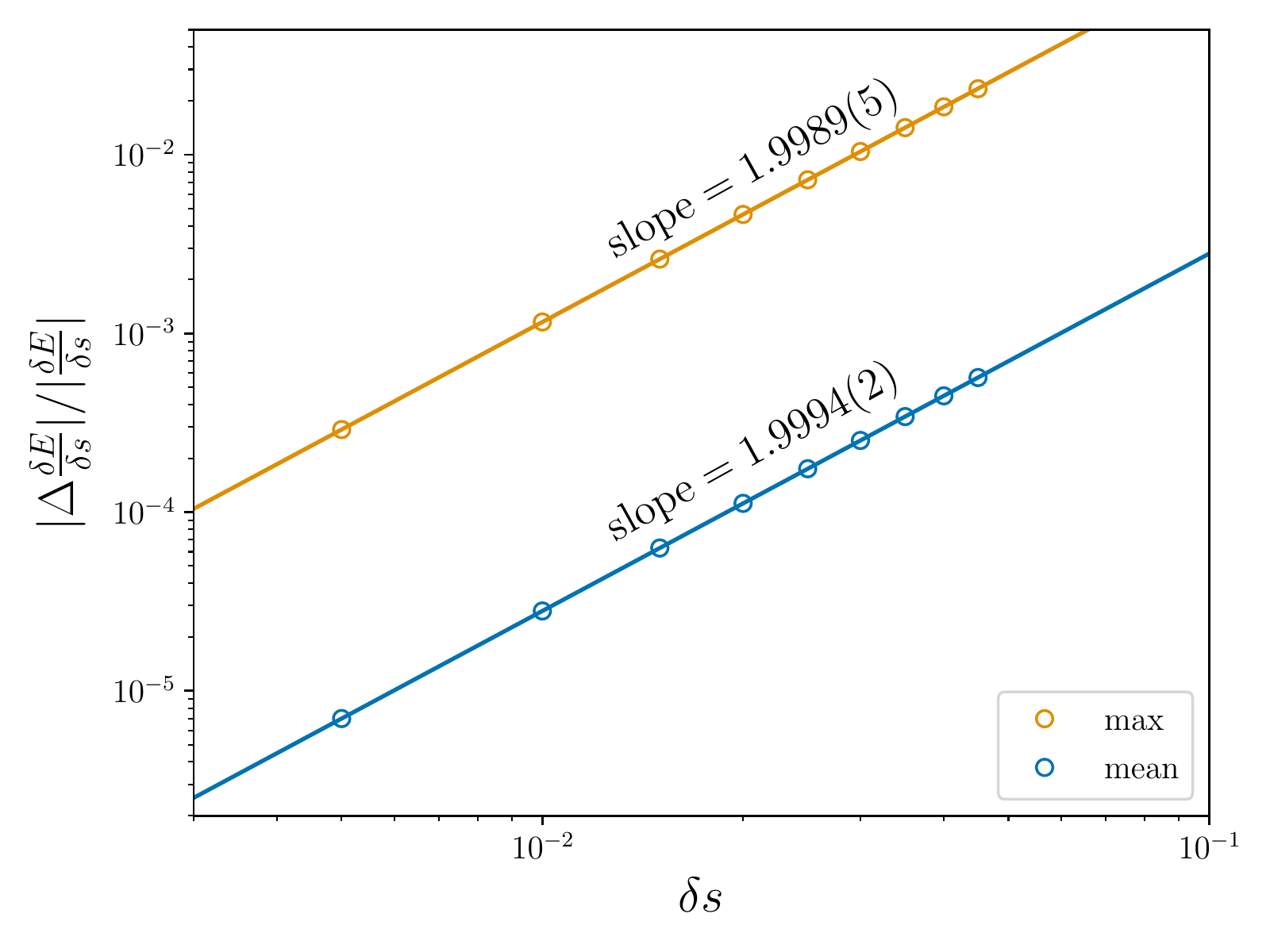}
  \caption{
Approximately quadratic convergence for Eq.~\eqref{eq:energy_derivitive} at $(\lb,\gamma)=(0.5,4)$. The number in brackets shown gives the fit error in the slope. The production run with default lattice spacing corresponds to $\delta s=0.01$. Here we have fixed $\delta z = 0.094$.
  }
  \label{fig:convergence_scalar}
\end{figure}

\vspace{4mm} At the parameter point $(\lb,\gamma)=(0.5,4)$, Fig.~\ref{fig:convergence_scalar} demonstrates that as the lattice spacing is decreased, $\delta z, \delta s \to 0_+$, the exact equality \eqref{eq:energy_derivitive} is approached quadratically, as expected for the leap-frog algorithm.

For the default lattice spacing choice (see Eq.~\eqref{eq:dz}), the maximum relative error in Eq.~\eqref{eq:energy_derivitive} occurs near the collision point and is approximately $0.1\%$, while the mean relative error is $0.003\%$.

The implementation of the discrete mode expansion in Appendix~\ref{appendix:fourier} was demonstrated to be orthogonal, at the level of machine precision.
In addition, it was shown to agree to high accuracy with the analytic result for a Gaussian blob.

The implementation of the numerical integrations determining the gravitational wave spectrum was compared with an independent implementation in {\tt Mathematica} using the inbuilt function {\tt NIntegrate}.
For a set of specific analytic field configurations, the two implementations were shown to agree to high accuracy, with the discrepancy approaching zero quadratically as the lattice spacing decreased, as expected for the trapezium rule.

\begin{figure}[t]
  \centering
    \includegraphics[width=0.48\textwidth]{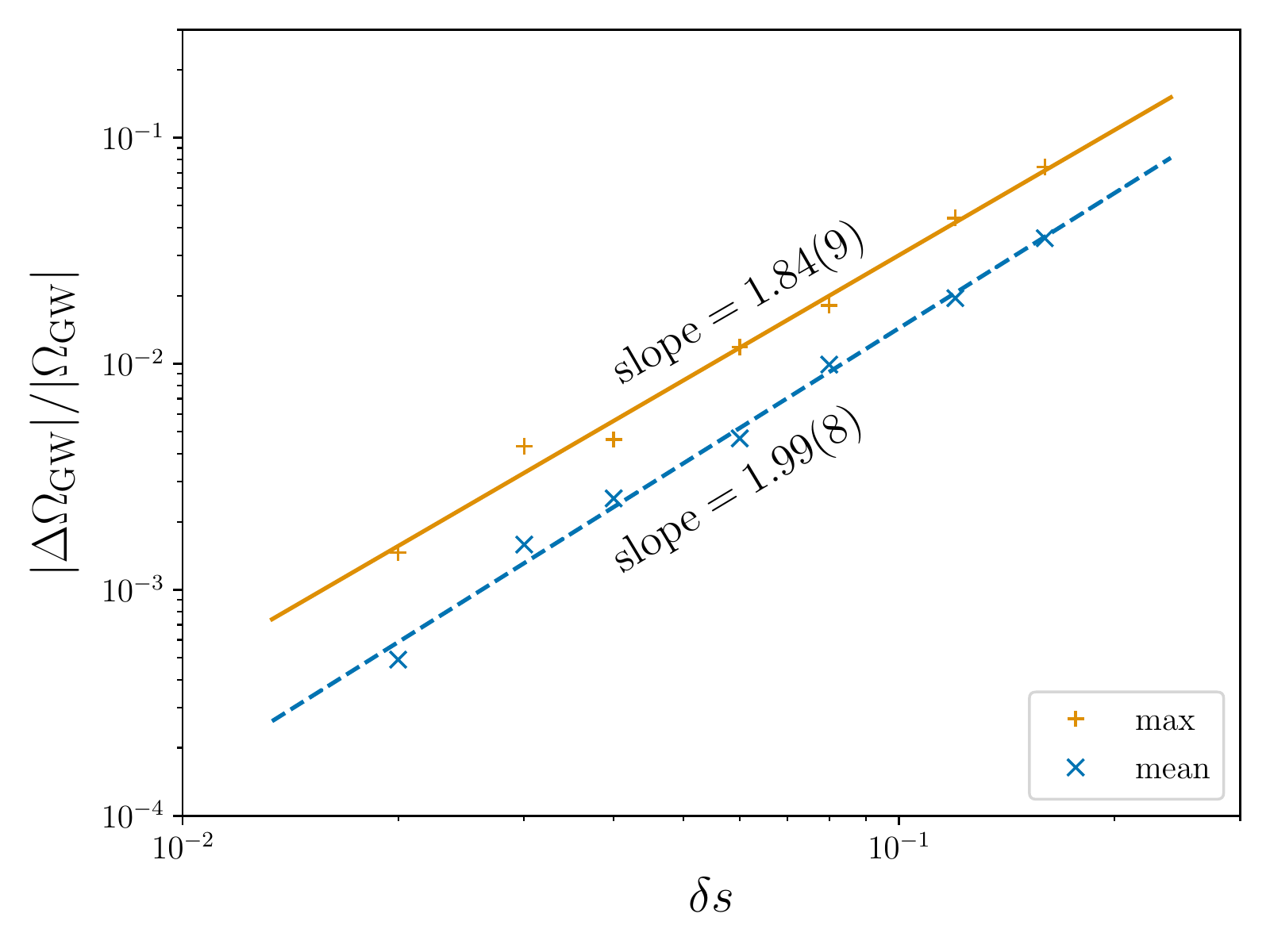}
  \caption{
  Approximately quadratic convergence for the GW spectrum, at $(\lb,\gamma)=(0.5,4)$, over the frequency range which is fitted. The number in brackets shown gives the fit error in the slope. The production run corresponds to $\delta s=0.05$. Here we have fixed the ratio $\delta z/\delta s = 2$.
  }
  \label{fig:convergence_gws_gamma_4}
\end{figure}

For one benchmark point, at $(\lb,\gamma)=(0.5,4)$, Fig.~\ref{fig:convergence_gws_gamma_4} demonstrates the approximately quadratic convergence for the gravitational wave spectrum as $\delta z$ and $\delta s$ are decreased towards the continuum limit.
Shown are the maximum and the mean absolute discrepancies between the gravitational wave spectrum in the run at a given $\delta s$ and that at $\delta s = 0.005$.
Only the fitted range, $\omega < \omega_{\rm cut}$, is included.
For the production run at this parameter point, which used $\delta s=0.05$, the fractional error is less than 1\%.

In addition to the aforementioned tests, which demonstrate the expected behaviour of the numerical codes as the continuum limit is approached, it is important to test the stability of our final results to changes in lattice spacing.
This is to test whether the values of $\delta z$ and $\delta s$ used, and enumerated in Appendix~\ref{appendix:table}, are small enough for our quantitative conclusions to be reliable.
For the scalar field simulation runs, the default lattice spacing $\delta z$ was chosen according to
\begin{align} \label{eq:dz}
\delta z &= \mr{min}\left(0.1,\frac{1}{10\ \gamma_{\rm alt}}\left(R_{\rm out}-R_{\rm in}\right)\right),
\end{align}
where $\gamma_{\rm alt}\approx \gamma$ is defined in
Eq.~\eqref{eq:gamma_d_Rin_Rout}.  This ensures that there are at least
ten lattice points across the bubble wall at the collision point.  
For some runs at $\lb=0.01$,
a smaller value of $\delta z$ was chosen.
The lattice spacing $\delta s$ was chosen to be smaller than $\delta
z$.  The complete list of all run parameters are enumerated in
Appendix~\ref{appendix:table}.  As the computation of the
gravitational wave signal is the most computationally intensive step,
for this step the field was down-sampled in the $s$-direction, with
only one in $N_{\delta s}$ points used.

\begin{figure}[t]
\centering
\setlength{\hfuzz}{1.1\columnwidth}
 \ttfamily
  \begin{subfigure}{0.48\textwidth}
    \includegraphics[width=\textwidth]{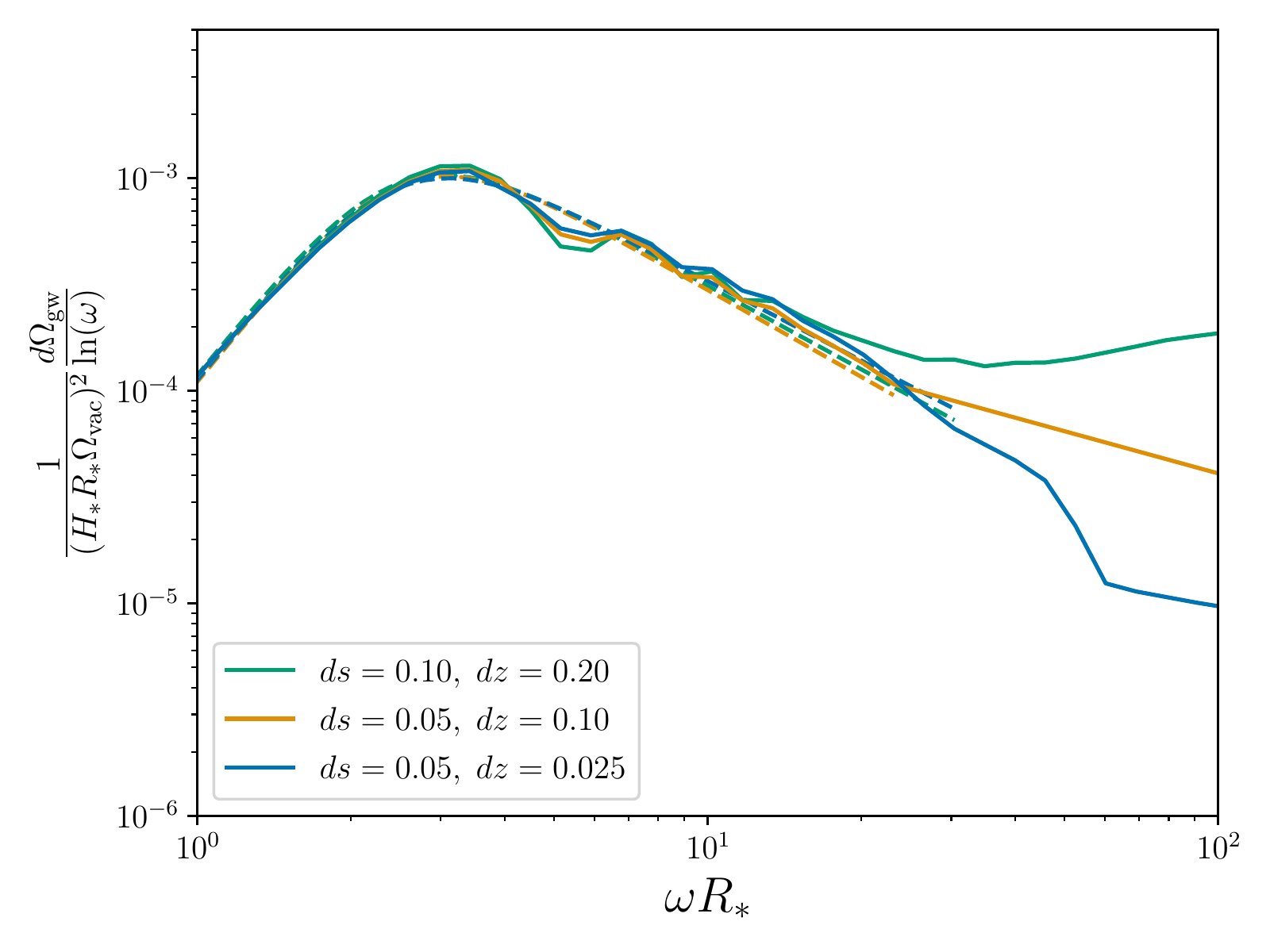}
    \caption{
      $(\lb,\gamma)=(0.84,16)$
    }
    \label{fig:convergence_gws_lambda_0.01_gamma_16}
  \end{subfigure}
    \begin{subfigure}{0.48\textwidth}
    \includegraphics[width=\textwidth]{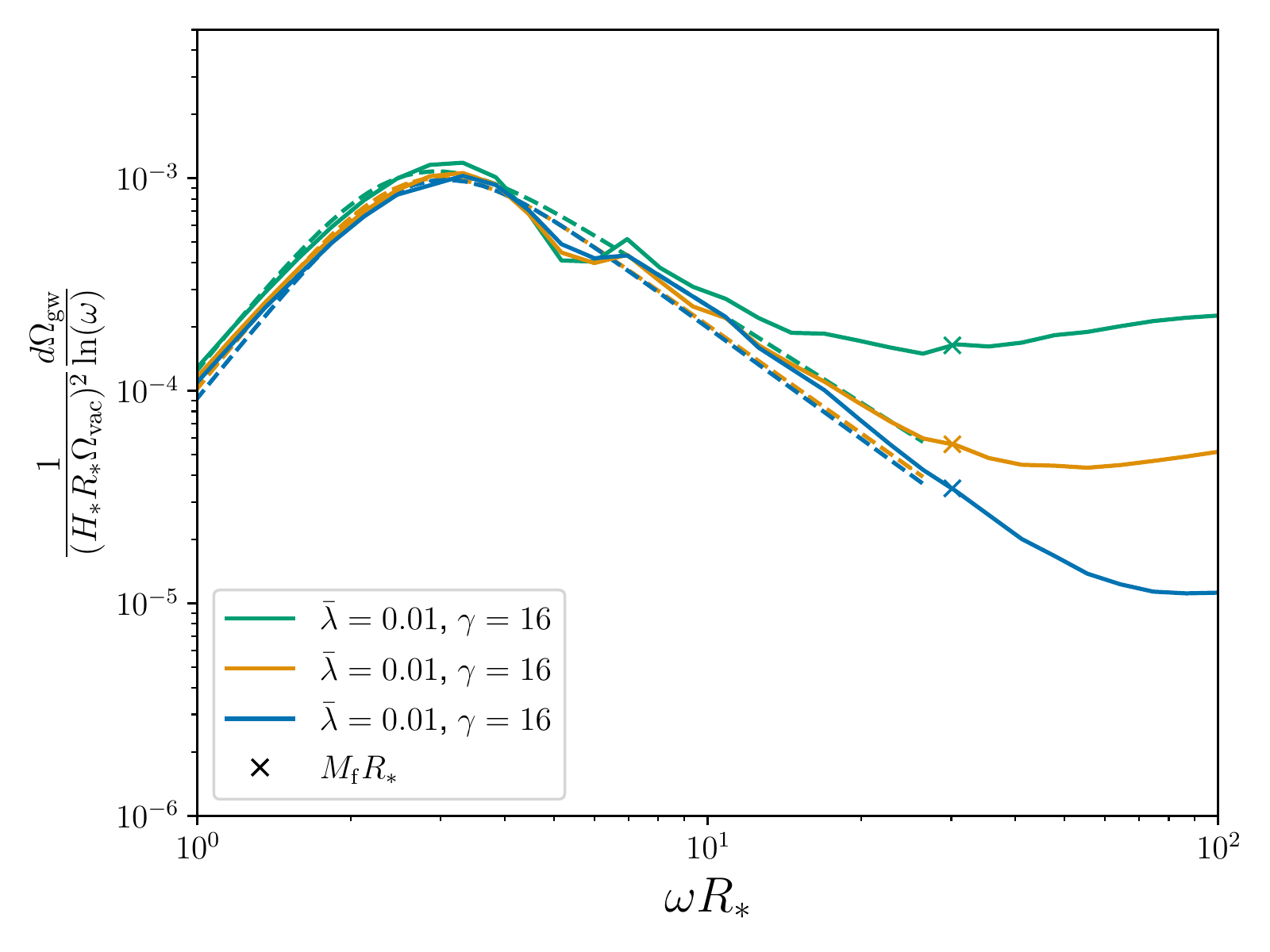}
      \caption{
      $(\lb,\gamma)=(0.01,16)$
      }
    \label{fig:convergence_gws_0.84_gamma_16}
  \end{subfigure}
  \caption{
	Lattice discretisation effects in the gravitational wave spectrum for the two runs with the largest hierarchies of scale, i.e.\ worst case scenarios.
	In each case the production runs correspond to those with smallest lattice spacing.
	Note that while the discrepancies in the spectrum are large towards $\omega R_*\sim 100$, the resulting effect on the fits is minimal.
	The mass scales which are not shown are greater than $100/R_*$.
  }
  \label{fig:convergence_gws_gamma_16}

\end{figure}

In Fig.~\ref{fig:convergence_gws_gamma_16}, we demonstrate the $\delta z$ and $\delta s$ dependence of the gravitational wave spectrum for two-bubble collisions with $\gamma=16$, one thin-wall with $\lb=0.84$ and one thick-wall with $\lb=0.01$.
In each case the lattice spacing used in the production run is compared to runs with two and four times larger lattice spacings.
The two parameter points are those with the largest hierarchies of scale, and hence where we expect the largest lattice discretisation errors.
A comparison of the discretisation errors at different parameter points bears this expectation out; the differences shown in Fig.~\ref{fig:convergence_gws_gamma_16} are larger than those at all other parameter points tested.
Nevertheless, significant disagreements between the spectra occur only for $\omega \gtrsim \omega_{\rm cut}$, i.e.\ frequencies that are not included in the fit.
\onecolumngrid

\twocolumngrid

\noindent Further, the fit is dominated by the region around the peak, so disagreement in the vicinity of $\omega\approx \omega_{\rm cut}$ has only a minor effect on the fit parameters.
Discrepancies between the fit parameters for the smaller two lattice
spacings are in the range 2--9\%, for those results shown in Fig.~\ref{fig:convergence_gws_gamma_16}.
This is comparable in magnitude with the fit error.
In addition, the largest discrepancies occur for the largest lattice spacings, suggesting convergence.

\section{Additional fit parameters} \label{appendix:extra_fits}

\begin{figure}[b]
  \centering
  \setlength{\hfuzz}{1.1\columnwidth}
\begin{minipage}{\textwidth}
 \ttfamily
  \begin{subfigure}{0.48\textwidth}
    \includegraphics[width=\textwidth]{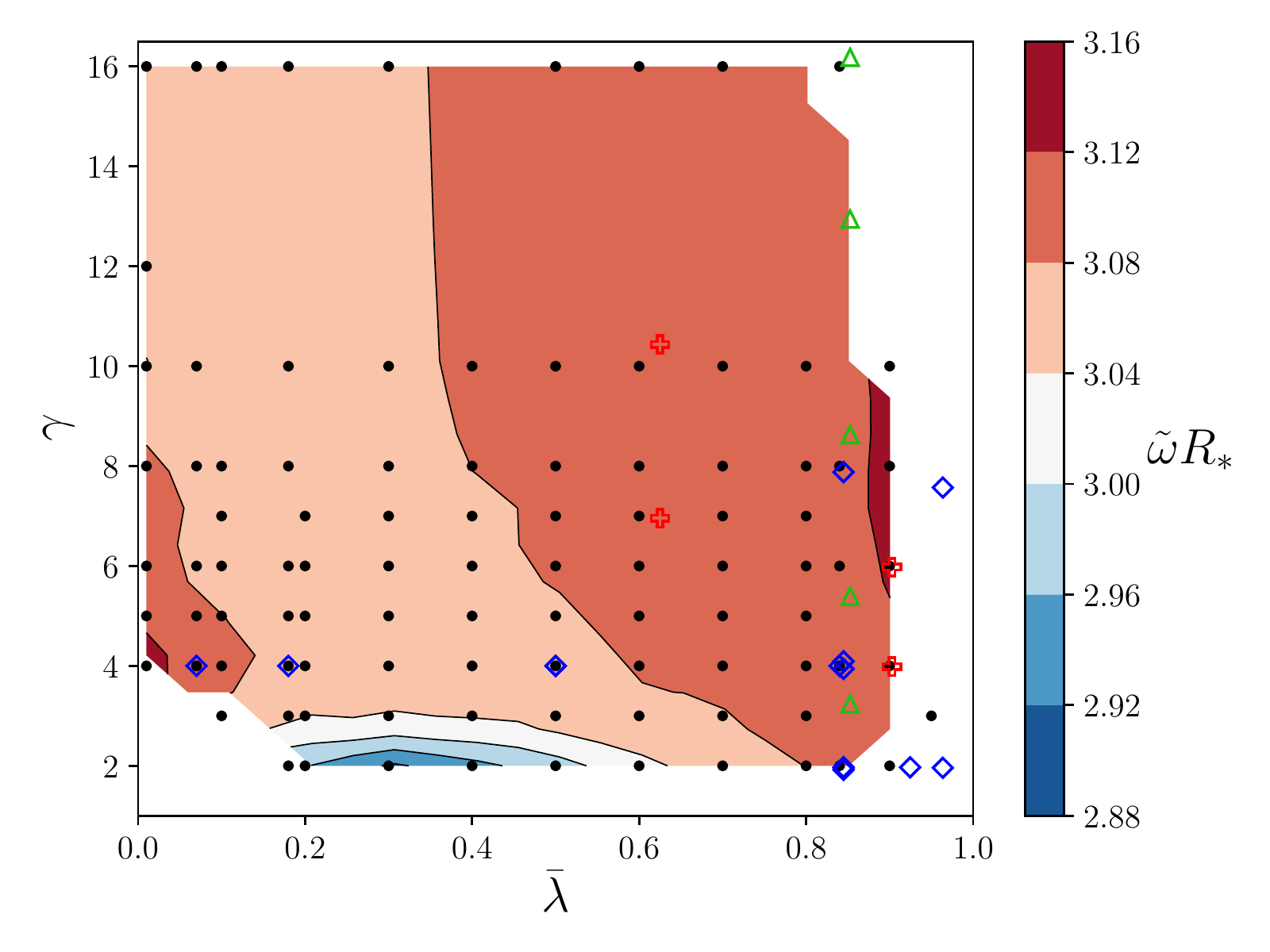}
    \caption{}
    \label{fig:k_fit}
  \end{subfigure}
    \begin{subfigure}{0.48\textwidth}
    \includegraphics[width=\textwidth]{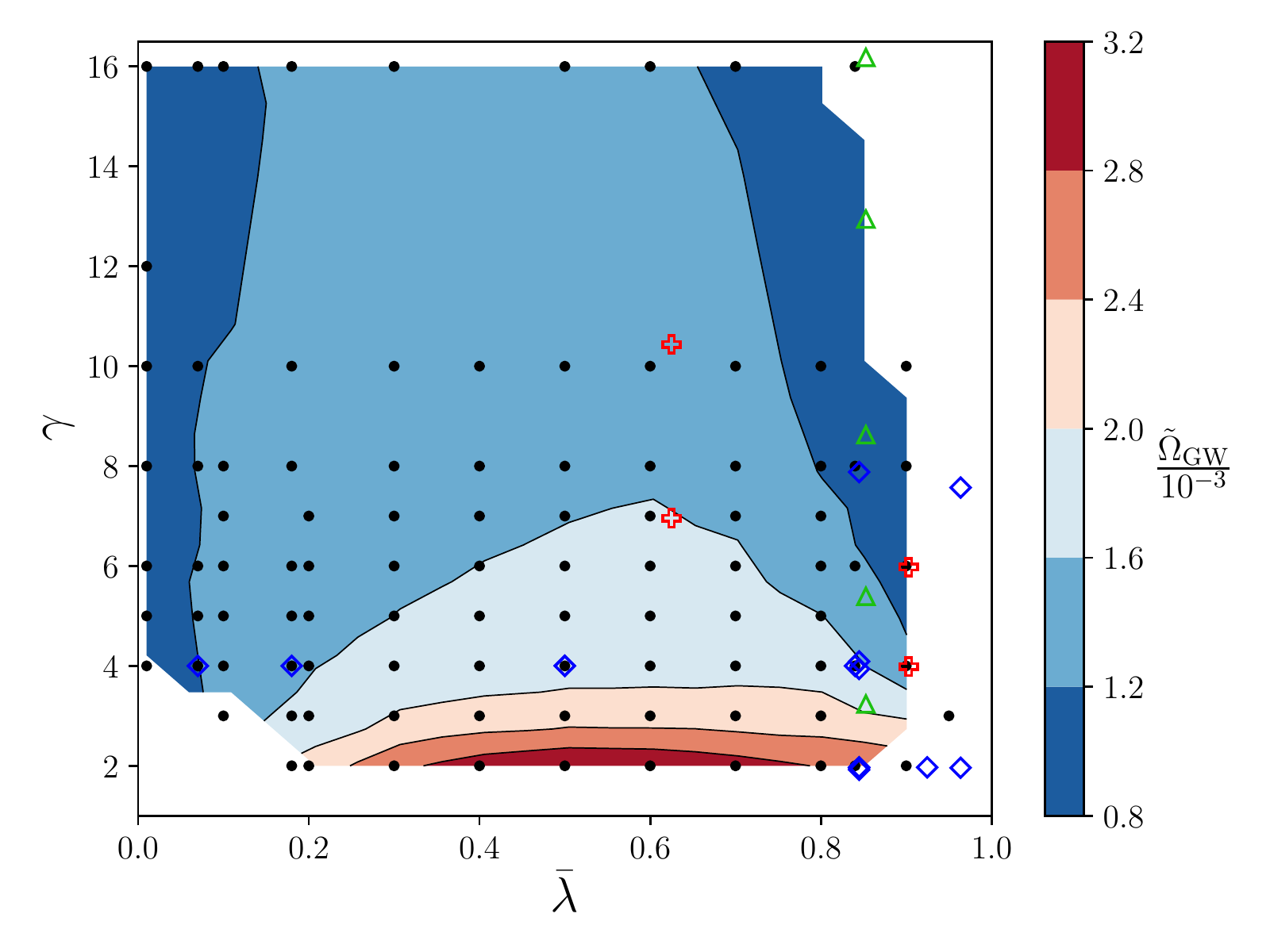}
    \caption{}
    \label{fig:Omega_fit}
  \end{subfigure}
  \caption{
	The other fit parameters in Eq.~\eqref{eq:fit_fn}, $\tilde{\omega}$ and $\tilde{\Omega}_{\rm GW}$, approximately equal to the peak position and amplitude respectively.
	Data points and contours are as Fig.~\ref{fig:b_contour}.
  }
  \label{fig:extra_fits}
  \end{minipage}
\end{figure}

For completeness, in Fig.~\ref{fig:extra_fits} we present the $\lb$ and $\gamma$ dependence of the other two fit parameters in Eq.~\eqref{eq:fit_fn},
$\tilde{\omega}$ and $\tilde{\Omega}_{\rm GW}$. Note that we fix $a=3$.
\onecolumngrid

\twocolumngrid

\clearpage
\newpage

\begin{widetext}
\section{Table of simulations} \label{appendix:table}
Note that this table only includes simulation runs used in the
preparation of the final results for this paper. See
Appendix~\ref{appendix:tests} for details of simulations carried out
as numerical tests.

\bigskip
\noindent

\begin{ruledtabular}
\begin{tabular}{D{.}{.}{1.2} D{.}{.}{2.0} |  D{.}{.}{2.2} D{.}{.}{2.2} D{.}{.}{2.2} | D{.}{.}{3.2} D{.}{.}{1.2} D{.}{.}{1.2} | c | D{.}{.}{1.3} c D{.}{.}{1.3} D{.}{.}{1.3} c D{.}{.}{1.3}
                D{.}{.}{1.3} c D{.}{.}{1.3}}
\multicolumn{2}{c|}{Parameters} & \multicolumn{3}{c|}{Bubble geometry} & \multicolumn{3}{c|}{Simulation} & \multicolumn{1}{c|}{Integration} & \multicolumn{9}{c}{Fitting results} \\
\hline
\multicolumn{1}{c}{$\overline{\lambda}$} & \multicolumn{1}{c}{$\gamma$} & \multicolumn{1}{c}{$R_0$} & \multicolumn{1}{c}{$R_\text{in}$} & \multicolumn{1}{c}{$R_\text{out}$} & \multicolumn{1}{c}{$d$} & \multicolumn{1}{c}{$\delta z$} & \multicolumn{1}{c}{$\delta s$} & \multicolumn{1}{c}{$N_\mathrm{\delta s}$} & \multicolumn{3}{c}{$\tilde{\Omega}_{\rm GW} \times 10^{3}$} & \multicolumn{3}{c}{$\tilde{\omega}R_*$} & \multicolumn{3}{c}{$b$} \\
\hline

0.01 & 2 & 19.98 & 12.88 & 29.31 & 79.92 & 0.10 & 0.01 & 5 & 0.518 & $\pm$ & 0.008 & 4.135 & $\pm$ & 0.161 & 0.612 & $\pm$ & 0.087\\ 
0.01 & 4 & 19.98 & 12.88 & 29.31 & 159.85 & 0.10 & 0.01 & 5 & 0.818 & $\pm$ & 0.018 & 3.142 & $\pm$ & 0.039 & 1.912 & $\pm$ & 0.125\\ 
0.01 & 5 & 19.98 & 12.88 & 29.31 & 199.81 & 0.10 & 0.01 & 5 & 0.922 & $\pm$ & 0.020 & 3.109 & $\pm$ & 0.039 & 1.863 & $\pm$ & 0.116\\ 
0.01 & 6 & 19.98 & 12.88 & 29.31 & 239.77 & 0.10 & 0.01 & 5 & 0.960 & $\pm$ & 0.023 & 3.088 & $\pm$ & 0.046 & 1.686 & $\pm$ & 0.103\\ 
0.01 & 8 & 19.98 & 12.88 & 29.31 & 319.70 & 0.05 & 0.01 & 5 & 0.991 & $\pm$ & 0.023 & 3.092 & $\pm$ & 0.045 & 1.650 & $\pm$ & 0.092\\ 
0.01 & 10 & 19.98 & 12.88 & 29.31 & 399.62 & 0.10 & 0.01 & 5 & 1.003 & $\pm$ & 0.025 & 3.038 & $\pm$ & 0.048 & 1.661 & $\pm$ & 0.096\\ 
0.01 & 12 & 19.98 & 12.88 & 29.31 & 479.55 & 0.05 & 0.01 & 5 & 1.020 & $\pm$ & 0.022 & 3.058 & $\pm$ & 0.043 & 1.679 & $\pm$ & 0.085\\ 
0.01 & 16 & 19.98 & 12.88 & 29.31 & 639.39 & 0.05 & 0.03 & 1 & 0.984 & $\pm$ & 0.028 & 3.068 & $\pm$ & 0.051 & 1.743 & $\pm$ & 0.118\\ 
0.07 & 2 & 7.99 & 5.20 & 11.67 & 31.95 & 0.10 & 0.01 & 5 & 1.087 & $\pm$ & 0.008 & 3.302 & $\pm$ & 0.054 & 1.124 & $\pm$ & 0.106\\ 
0.07 & 3 & 7.99 & 5.20 & 11.67 & 47.92 & 0.10 & 0.01 & 5 & 1.198 & $\pm$ & 0.017 & 3.097 & $\pm$ & 0.024 & 1.920 & $\pm$ & 0.100\\ 
0.07 & 4 & 7.99 & 5.20 & 11.67 & 63.89 & 0.10 & 0.01 & 5 & 1.227 & $\pm$ & 0.016 & 3.100 & $\pm$ & 0.025 & 1.595 & $\pm$ & 0.065\\ 
0.07 & 5 & 7.99 & 5.20 & 11.67 & 79.86 & 0.10 & 0.01 & 5 & 1.248 & $\pm$ & 0.027 & 3.087 & $\pm$ & 0.043 & 1.525 & $\pm$ & 0.092\\ 
0.07 & 6 & 7.99 & 5.20 & 11.67 & 95.84 & 0.10 & 0.01 & 5 & 1.251 & $\pm$ & 0.028 & 3.073 & $\pm$ & 0.045 & 1.481 & $\pm$ & 0.087\\ 
0.07 & 8 & 7.99 & 5.20 & 11.67 & 127.78 & 0.09 & 0.01 & 5 & 1.234 & $\pm$ & 0.028 & 3.063 & $\pm$ & 0.046 & 1.476 & $\pm$ & 0.081\\ 
0.07 & 10 & 7.99 & 5.20 & 11.67 & 159.73 & 0.07 & 0.01 & 5 & 1.204 & $\pm$ & 0.026 & 3.060 & $\pm$ & 0.045 & 1.485 & $\pm$ & 0.074\\ 
0.07 & 16 & 7.99 & 5.20 & 11.67 & 255.56 & 0.04 & 0.01 & 5 & 1.147 & $\pm$ & 0.024 & 3.063 & $\pm$ & 0.044 & 1.556 & $\pm$ & 0.074\\ 
0.10 & 2 & 6.94 & 4.55 & 10.09 & 27.77 & 0.10 & 0.02 & 5 & 1.353 & $\pm$ & 0.012 & 3.142 & $\pm$ & 0.037 & 1.344 & $\pm$ & 0.113\\ 
0.10 & 3 & 6.94 & 4.55 & 10.09 & 41.66 & 0.10 & 0.02 & 5 & 1.355 & $\pm$ & 0.020 & 3.075 & $\pm$ & 0.025 & 1.813 & $\pm$ & 0.092\\ 
0.10 & 4 & 6.94 & 4.55 & 10.09 & 55.55 & 0.10 & 0.02 & 5 & 1.327 & $\pm$ & 0.029 & 3.092 & $\pm$ & 0.042 & 1.525 & $\pm$ & 0.100\\ 
0.10 & 5 & 6.94 & 4.55 & 10.09 & 69.43 & 0.10 & 0.02 & 5 & 1.326 & $\pm$ & 0.031 & 3.080 & $\pm$ & 0.046 & 1.487 & $\pm$ & 0.093\\ 
0.10 & 6 & 6.94 & 4.55 & 10.09 & 83.32 & 0.10 & 0.02 & 5 & 1.313 & $\pm$ & 0.031 & 3.073 & $\pm$ & 0.048 & 1.474 & $\pm$ & 0.089\\ 
0.10 & 7 & 6.94 & 4.55 & 10.09 & 97.20 & 0.08 & 0.02 & 5 & 1.299 & $\pm$ & 0.031 & 3.069 & $\pm$ & 0.048 & 1.474 & $\pm$ & 0.085\\ 
0.10 & 8 & 6.94 & 4.55 & 10.09 & 111.09 & 0.07 & 0.01 & 5 & 1.277 & $\pm$ & 0.027 & 3.066 & $\pm$ & 0.044 & 1.457 & $\pm$ & 0.075\\ 
0.10 & 16 & 6.94 & 4.55 & 10.09 & 222.18 & 0.04 & 0.03 & 1 & 1.168 & $\pm$ & 0.022 & 3.070 & $\pm$ & 0.040 & 1.521 & $\pm$ & 0.066\\ 
0.18 & 2 & 5.79 & 3.81 & 8.26 & 23.18 & 0.10 & 0.01 & 5 & 2.003 & $\pm$ & 0.025 & 2.971 & $\pm$ & 0.023 & 1.652 & $\pm$ & 0.106\\ 
0.18 & 3 & 5.79 & 3.81 & 8.26 & 34.76 & 0.10 & 0.01 & 5 & 1.656 & $\pm$ & 0.032 & 3.057 & $\pm$ & 0.037 & 1.452 & $\pm$ & 0.097\\ 
0.18 & 4 & 5.79 & 3.81 & 8.26 & 46.35 & 0.10 & 0.01 & 5 & 1.532 & $\pm$ & 0.019 & 3.073 & $\pm$ & 0.025 & 1.394 & $\pm$ & 0.052\\ 
0.18 & 5 & 5.79 & 3.81 & 8.26 & 57.94 & 0.09 & 0.01 & 5 & 1.464 & $\pm$ & 0.031 & 3.074 & $\pm$ & 0.043 & 1.355 & $\pm$ & 0.075\\ 
0.18 & 6 & 5.79 & 3.81 & 8.26 & 69.53 & 0.08 & 0.01 & 5 & 1.421 & $\pm$ & 0.029 & 3.073 & $\pm$ & 0.043 & 1.366 & $\pm$ & 0.070\\ 
0.18 & 8 & 5.79 & 3.81 & 8.26 & 92.70 & 0.06 & 0.01 & 5 & 1.359 & $\pm$ & 0.027 & 3.071 & $\pm$ & 0.042 & 1.405 & $\pm$ & 0.066\\ 
0.18 & 10 & 5.79 & 3.81 & 8.26 & 115.88 & 0.05 & 0.01 & 5 & 1.316 & $\pm$ & 0.026 & 3.070 & $\pm$ & 0.041 & 1.440 & $\pm$ & 0.065\\ 
0.18 & 16 & 5.79 & 3.81 & 8.26 & 185.40 & 0.03 & 0.01 & 5 & 1.231 & $\pm$ & 0.025 & 3.072 & $\pm$ & 0.043 & 1.508 & $\pm$ & 0.069\\ 
0.20 & 2 & 5.65 & 3.75 & 8.03 & 22.61 & 0.10 & 0.02 & 5 & 2.124 & $\pm$ & 0.026 & 2.963 & $\pm$ & 0.023 & 1.637 & $\pm$ & 0.106\\ 
0.20 & 3 & 5.65 & 3.75 & 8.03 & 33.92 & 0.10 & 0.02 & 5 & 1.723 & $\pm$ & 0.034 & 3.049 & $\pm$ & 0.037 & 1.449 & $\pm$ & 0.098\\ 
0.20 & 4 & 5.65 & 3.75 & 8.03 & 45.22 & 0.10 & 0.02 & 5 & 1.566 & $\pm$ & 0.033 & 3.075 & $\pm$ & 0.043 & 1.365 & $\pm$ & 0.084\\ 
0.20 & 5 & 5.65 & 3.75 & 8.03 & 56.53 & 0.09 & 0.02 & 5 & 1.495 & $\pm$ & 0.033 & 3.074 & $\pm$ & 0.045 & 1.357 & $\pm$ & 0.079\\ 
0.20 & 6 & 5.65 & 3.75 & 8.03 & 67.83 & 0.07 & 0.02 & 5 & 1.449 & $\pm$ & 0.032 & 3.071 & $\pm$ & 0.046 & 1.373 & $\pm$ & 0.075\\ 
0.20 & 7 & 5.65 & 3.75 & 8.03 & 79.14 & 0.06 & 0.02 & 5 & 1.411 & $\pm$ & 0.031 & 3.073 & $\pm$ & 0.046 & 1.397 & $\pm$ & 0.073\\ 
0.30 & 2 & 5.40 & 3.65 & 7.53 & 21.59 & 0.10 & 0.02 & 5 & 2.695 & $\pm$ & 0.030 & 2.912 & $\pm$ & 0.019 & 1.602 & $\pm$ & 0.073\\ 
0.30 & 3 & 5.40 & 3.65 & 7.53 & 32.39 & 0.10 & 0.02 & 5 & 1.953 & $\pm$ & 0.041 & 3.054 & $\pm$ & 0.042 & 1.246 & $\pm$ & 0.085\\ 
0.30 & 4 & 5.40 & 3.65 & 7.53 & 43.19 & 0.10 & 0.02 & 5 & 1.715 & $\pm$ & 0.036 & 3.070 & $\pm$ & 0.045 & 1.236 & $\pm$ & 0.073\\ 
0.30 & 5 & 5.40 & 3.65 & 7.53 & 53.98 & 0.08 & 0.02 & 5 & 1.602 & $\pm$ & 0.033 & 3.076 & $\pm$ & 0.045 & 1.275 & $\pm$ & 0.068\\ 
0.30 & 6 & 5.40 & 3.65 & 7.53 & 64.78 & 0.07 & 0.02 & 5 & 1.537 & $\pm$ & 0.031 & 3.074 & $\pm$ & 0.044 & 1.312 & $\pm$ & 0.065\\ 
0.30 & 7 & 5.40 & 3.65 & 7.53 & 75.58 & 0.06 & 0.02 & 5 & 1.492 & $\pm$ & 0.030 & 3.073 & $\pm$ & 0.044 & 1.346 & $\pm$ & 0.064\\ 
0.30 & 8 & 5.40 & 3.65 & 7.53 & 86.37 & 0.05 & 0.01 & 5 & 1.451 & $\pm$ & 0.027 & 3.074 & $\pm$ & 0.040 & 1.356 & $\pm$ & 0.059\\ 
0.30 & 10 & 5.40 & 3.65 & 7.53 & 107.97 & 0.04 & 0.01 & 5 & 1.395 & $\pm$ & 0.026 & 3.076 & $\pm$ & 0.040 & 1.397 & $\pm$ & 0.058\\ 
0.30 & 16 & 5.40 & 3.65 & 7.53 & 172.75 & 0.03 & 0.01 & 5 & 1.286 & $\pm$ & 0.026 & 3.077 & $\pm$ & 0.043 & 1.456 & $\pm$ & 0.066\\ 
0.40 & 2 & 5.60 & 3.89 & 7.61 & 22.40 & 0.10 & 0.02 & 5 & 3.001 & $\pm$ & 0.055 & 2.947 & $\pm$ & 0.036 & 1.241 & $\pm$ & 0.090\\ 
0.40 & 3 & 5.60 & 3.89 & 7.61 & 33.61 & 0.10 & 0.02 & 5 & 2.093 & $\pm$ & 0.044 & 3.058 & $\pm$ & 0.045 & 1.121 & $\pm$ & 0.072\\ 
0.40 & 4 & 5.60 & 3.89 & 7.61 & 44.81 & 0.10 & 0.02 & 5 & 1.809 & $\pm$ & 0.037 & 3.074 & $\pm$ & 0.045 & 1.185 & $\pm$ & 0.066\\ 
0.40 & 5 & 5.60 & 3.89 & 7.61 & 56.01 & 0.08 & 0.02 & 5 & 1.678 & $\pm$ & 0.033 & 3.074 & $\pm$ & 0.043 & 1.230 & $\pm$ & 0.059\\ 
0.40 & 6 & 5.60 & 3.89 & 7.61 & 67.21 & 0.06 & 0.02 & 5 & 1.599 & $\pm$ & 0.031 & 3.077 & $\pm$ & 0.042 & 1.280 & $\pm$ & 0.058\\ 
0.40 & 7 & 5.60 & 3.89 & 7.61 & 78.41 & 0.05 & 0.02 & 5 & 1.549 & $\pm$ & 0.030 & 3.076 & $\pm$ & 0.043 & 1.318 & $\pm$ & 0.059\\ 
0.40 & 8 & 5.60 & 3.89 & 7.61 & 89.61 & 0.05 & 0.01 & 5 & 1.504 & $\pm$ & 0.027 & 3.080 & $\pm$ & 0.039 & 1.335 & $\pm$ & 0.054\\ 
0.40 & 10 & 5.60 & 3.89 & 7.61 & 112.02 & 0.04 & 0.01 & 5 & 1.445 & $\pm$ & 0.026 & 3.083 & $\pm$ & 0.039 & 1.371 & $\pm$ & 0.055\\ 
0.50 & 2 & 6.21 & 4.45 & 8.16 & 24.83 & 0.10 & 0.01 & 5 & 3.166 & $\pm$ & 0.065 & 2.983 & $\pm$ & 0.044 & 1.024 & $\pm$ & 0.078\\ 
0.50 & 3 & 6.21 & 4.45 & 8.16 & 37.25 & 0.10 & 0.01 & 5 & 2.158 & $\pm$ & 0.044 & 3.066 & $\pm$ & 0.046 & 1.051 & $\pm$ & 0.060\\ 
0.50 & 4 & 6.21 & 4.45 & 8.16 & 49.66 & 0.09 & 0.01 & 5 & 1.866 & $\pm$ & 0.023 & 3.073 & $\pm$ & 0.027 & 1.156 & $\pm$ & 0.037\\ 
0.50 & 5 & 6.21 & 4.45 & 8.16 & 62.08 & 0.08 & 0.01 & 5 & 1.725 & $\pm$ & 0.032 & 3.077 & $\pm$ & 0.041 & 1.208 & $\pm$ & 0.053\\ 
0.50 & 6 & 6.21 & 4.45 & 8.16 & 74.50 & 0.06 & 0.01 & 5 & 1.642 & $\pm$ & 0.029 & 3.082 & $\pm$ & 0.040 & 1.259 & $\pm$ & 0.051\\ 
0.50 & 7 & 6.21 & 4.45 & 8.16 & 86.91 & 0.05 & 0.02 & 5 & 1.592 & $\pm$ & 0.030 & 3.082 & $\pm$ & 0.041 & 1.305 & $\pm$ & 0.055\\ 
0.50 & 8 & 6.21 & 4.45 & 8.16 & 99.33 & 0.05 & 0.01 & 5 & 1.547 & $\pm$ & 0.027 & 3.088 & $\pm$ & 0.039 & 1.319 & $\pm$ & 0.052\\ 
0.50 & 10 & 6.21 & 4.45 & 8.16 & 124.16 & 0.04 & 0.01 & 5 & 1.470 & $\pm$ & 0.027 & 3.092 & $\pm$ & 0.040 & 1.332 & $\pm$ & 0.054\\ 
0.50 & 16 & 6.21 & 4.45 & 8.16 & 198.65 & 0.02 & 0.01 & 5 & 1.296 & $\pm$ & 0.026 & 3.089 & $\pm$ & 0.042 & 1.442 & $\pm$ & 0.063\\ 
0.60 & 2 & 7.29 & 5.47 & 9.24 & 29.16 & 0.10 & 0.02 & 5 & 3.129 & $\pm$ & 0.064 & 3.029 & $\pm$ & 0.048 & 0.892 & $\pm$ & 0.060\\ 
0.60 & 3 & 7.29 & 5.47 & 9.24 & 43.74 & 0.10 & 0.02 & 5 & 2.155 & $\pm$ & 0.042 & 3.077 & $\pm$ & 0.045 & 1.029 & $\pm$ & 0.053\\ 
0.60 & 4 & 7.29 & 5.47 & 9.24 & 58.32 & 0.09 & 0.02 & 5 & 1.880 & $\pm$ & 0.035 & 3.081 & $\pm$ & 0.043 & 1.140 & $\pm$ & 0.051\\ 
0.60 & 5 & 7.29 & 5.47 & 9.24 & 72.90 & 0.08 & 0.02 & 5 & 1.754 & $\pm$ & 0.032 & 3.086 & $\pm$ & 0.041 & 1.215 & $\pm$ & 0.050\\ 
0.60 & 6 & 7.29 & 5.47 & 9.24 & 87.48 & 0.06 & 0.02 & 5 & 1.681 & $\pm$ & 0.030 & 3.091 & $\pm$ & 0.040 & 1.264 & $\pm$ & 0.051\\ 
0.60 & 7 & 7.29 & 5.47 & 9.24 & 102.06 & 0.05 & 0.02 & 5 & 1.626 & $\pm$ & 0.030 & 3.095 & $\pm$ & 0.042 & 1.284 & $\pm$ & 0.054\\ 
0.60 & 8 & 7.29 & 5.47 & 9.24 & 116.64 & 0.05 & 0.01 & 5 & 1.557 & $\pm$ & 0.028 & 3.100 & $\pm$ & 0.041 & 1.279 & $\pm$ & 0.051\\ 
0.60 & 10 & 7.29 & 5.47 & 9.24 & 145.79 & 0.04 & 0.01 & 5 & 1.450 & $\pm$ & 0.026 & 3.098 & $\pm$ & 0.039 & 1.346 & $\pm$ & 0.053\\ 
0.60 & 16 & 7.29 & 5.47 & 9.24 & 233.27 & 0.02 & 0.01 & 5 & 1.237 & $\pm$ & 0.024 & 3.117 & $\pm$ & 0.042 & 1.447 & $\pm$ & 0.061\\ 
0.70 & 2 & 9.25 & 7.35 & 11.21 & 37.01 & 0.10 & 0.02 & 5 & 2.970 & $\pm$ & 0.059 & 3.062 & $\pm$ & 0.049 & 0.855 & $\pm$ & 0.050\\ 
0.70 & 3 & 9.25 & 7.35 & 11.21 & 55.51 & 0.10 & 0.02 & 5 & 2.139 & $\pm$ & 0.039 & 3.080 & $\pm$ & 0.043 & 1.042 & $\pm$ & 0.046\\ 
0.70 & 4 & 9.25 & 7.35 & 11.21 & 74.02 & 0.10 & 0.02 & 5 & 1.901 & $\pm$ & 0.034 & 3.088 & $\pm$ & 0.041 & 1.157 & $\pm$ & 0.047\\ 
0.70 & 5 & 9.25 & 7.35 & 11.21 & 92.52 & 0.08 & 0.02 & 5 & 1.787 & $\pm$ & 0.032 & 3.099 & $\pm$ & 0.041 & 1.213 & $\pm$ & 0.049\\ 
0.70 & 6 & 9.25 & 7.35 & 11.21 & 111.03 & 0.06 & 0.02 & 5 & 1.665 & $\pm$ & 0.031 & 3.108 & $\pm$ & 0.043 & 1.218 & $\pm$ & 0.052\\ 
0.70 & 7 & 9.25 & 7.35 & 11.21 & 129.53 & 0.06 & 0.02 & 5 & 1.554 & $\pm$ & 0.029 & 3.102 & $\pm$ & 0.042 & 1.281 & $\pm$ & 0.054\\ 
0.70 & 8 & 9.25 & 7.35 & 11.21 & 148.04 & 0.05 & 0.01 & 5 & 1.453 & $\pm$ & 0.025 & 3.106 & $\pm$ & 0.038 & 1.345 & $\pm$ & 0.052\\ 
0.70 & 10 & 9.25 & 7.35 & 11.21 & 185.05 & 0.04 & 0.01 & 5 & 1.293 & $\pm$ & 0.023 & 3.103 & $\pm$ & 0.039 & 1.374 & $\pm$ & 0.054\\ 
0.70 & 16 & 9.25 & 7.35 & 11.21 & 296.07 & 0.02 & 0.01 & 5 & 1.170 & $\pm$ & 0.023 & 3.105 & $\pm$ & 0.043 & 1.412 & $\pm$ & 0.063\\ 
0.80 & 2 & 13.17 & 11.19 & 15.15 & 52.68 & 0.10 & 0.02 & 5 & 2.774 & $\pm$ & 0.051 & 3.081 & $\pm$ & 0.047 & 0.872 & $\pm$ & 0.041\\ 
0.80 & 3 & 13.17 & 11.19 & 15.15 & 79.02 & 0.10 & 0.02 & 5 & 2.129 & $\pm$ & 0.037 & 3.095 & $\pm$ & 0.042 & 1.077 & $\pm$ & 0.043\\ 
0.80 & 4 & 13.17 & 11.19 & 15.15 & 105.36 & 0.10 & 0.01 & 5 & 1.870 & $\pm$ & 0.024 & 3.118 & $\pm$ & 0.030 & 1.104 & $\pm$ & 0.034\\ 
0.80 & 5 & 13.17 & 11.19 & 15.15 & 131.70 & 0.08 & 0.02 & 5 & 1.610 & $\pm$ & 0.028 & 3.110 & $\pm$ & 0.039 & 1.261 & $\pm$ & 0.049\\ 
0.80 & 6 & 13.17 & 11.19 & 15.15 & 158.04 & 0.07 & 0.02 & 5 & 1.380 & $\pm$ & 0.025 & 3.109 & $\pm$ & 0.040 & 1.292 & $\pm$ & 0.052\\ 
0.80 & 7 & 13.17 & 11.19 & 15.15 & 184.38 & 0.06 & 0.02 & 5 & 1.285 & $\pm$ & 0.024 & 3.107 & $\pm$ & 0.042 & 1.318 & $\pm$ & 0.055\\ 
0.80 & 8 & 13.17 & 11.19 & 15.15 & 210.72 & 0.05 & 0.01 & 5 & 1.175 & $\pm$ & 0.022 & 3.107 & $\pm$ & 0.041 & 1.283 & $\pm$ & 0.052\\ 
0.80 & 10 & 13.17 & 11.19 & 15.15 & 263.39 & 0.04 & 0.01 & 5 & 1.124 & $\pm$ & 0.021 & 3.108 & $\pm$ & 0.043 & 1.298 & $\pm$ & 0.055\\ 
0.84 & 2 & 16.06 & 14.07 & 18.02 & 64.23 & 0.10 & 0.01 & 5 & 2.719 & $\pm$ & 0.049 & 3.083 & $\pm$ & 0.046 & 0.899 & $\pm$ & 0.039\\ 
0.84 & 4 & 16.06 & 14.07 & 18.02 & 128.46 & 0.10 & 0.01 & 5 & 1.721 & $\pm$ & 0.020 & 3.109 & $\pm$ & 0.027 & 1.192 & $\pm$ & 0.032\\ 
0.84 & 6 & 16.06 & 14.07 & 18.02 & 192.69 & 0.07 & 0.01 & 5 & 1.211 & $\pm$ & 0.022 & 3.110 & $\pm$ & 0.041 & 1.236 & $\pm$ & 0.050\\ 
0.84 & 8 & 16.06 & 14.07 & 18.02 & 256.92 & 0.05 & 0.01 & 5 & 1.114 & $\pm$ & 0.021 & 3.111 & $\pm$ & 0.044 & 1.246 & $\pm$ & 0.053\\ 
0.84 & 16 & 16.06 & 14.07 & 18.02 & 513.84 & 0.02 & 0.01 & 5 & 1.001 & $\pm$ & 0.025 & 3.100 & $\pm$ & 0.057 & 1.246 & $\pm$ & 0.067\\ 
0.90 & 2 & 24.38 & 22.37 & 26.39 & 97.54 & 0.10 & 0.02 & 5 & 2.654 & $\pm$ & 0.049 & 3.101 & $\pm$ & 0.047 & 0.930 & $\pm$ & 0.040\\ 
0.90 & 4 & 24.38 & 22.37 & 26.39 & 195.08 & 0.10 & 0.01 & 5 & 1.273 & $\pm$ & 0.028 & 3.116 & $\pm$ & 0.046 & 1.196 & $\pm$ & 0.048\\ 
0.90 & 6 & 24.38 & 22.37 & 26.39 & 292.62 & 0.07 & 0.02 & 5 & 1.043 & $\pm$ & 0.021 & 3.122 & $\pm$ & 0.048 & 1.156 & $\pm$ & 0.053\\ 
0.90 & 8 & 24.38 & 22.37 & 26.39 & 390.15 & 0.05 & 0.01 & 5 & 1.026 & $\pm$ & 0.020 & 3.126 & $\pm$ & 0.046 & 1.186 & $\pm$ & 0.053\\ 
0.90 & 10 & 24.38 & 22.37 & 26.39 & 487.69 & 0.04 & 0.01 & 5 & 1.001 & $\pm$ & 0.021 & 3.124 & $\pm$ & 0.048 & 1.198 & $\pm$ & 0.056\\ 
0.95 & 3 & 45.96 & 43.92 & 48.01 & 275.79 & 0.10 & 0.01 & 5 & 1.207 & $\pm$ & 0.022 & 3.089 & $\pm$ & 0.041 & 1.211 & $\pm$ & 0.048\\ 

\end{tabular}
\end{ruledtabular}

\end{widetext}

\bibliography{refs}

\end{document}